\documentclass[useAMS, usenatbib, usegraphicx]{mn2e}
\usepackage{times}
\usepackage{amssymb,amsmath}
\usepackage{url}
\usepackage{graphicx}
\usepackage{subfigure}
\usepackage{float}

\def\ec{$\eta$~Car}
\def\eca{$\eta_{\mathrm{A}}$}
\def\ecb{$\eta_{\mathrm{B}}$}

\def\hst{{\it HST}}

\newcommand{\Lsun}{\hbox{$L_\odot$}}
\newcommand{\Msun}{\hbox{$M_\odot$}}

\newcommand{\altion}[2]{\textup{#1}\,\textsc{#2}}

\title[Constraining the 3-D Orientation of $\eta$ Car's Orbit]{Constraining the Absolute Orientation of $\boldeta$ Carinae's Binary Orbit: A 3-D Dynamical Model for the Broad [\altion{Fe}{iii}] Emission\thanks{Based on observations made with the NASA/ESA \emph{Hubble Space Telescope}. Support for programs 7302, 8036, 8483, 8619, 9083, 9337, 9420, 9973, 10957 and 11273 was provided by NASA directly to the STIS Science Team and through grants from the Space Telescope Science Institute, which is operated by the Association of Universities for Research in Astronomy, Inc., under NASA contract NAS~5-26555.}}

\author[Madura et al.]{T.~I. Madura$^{1}$\thanks{Email: tmadura@mpifr-bonn.mpg.de},
T.~R. Gull$^{2}$,
S.~P. Owocki$^{3}$,
J.~H. Groh$^{1}$,
A.~T. Okazaki$^{4}$,
and C.~M.~P. Russell$^{3}$ \\
$^{1}$ Max-Planck-Institut f\"{u}r Radioastronomie, Auf dem H\"{u}gel 69, D-53121 Bonn, Germany\\
$^{2}$ Astrophysics Science Division, Code 667, NASA Goddard Space Flight Center, Greenbelt, MD 20771, USA\\
$^{3}$ Bartol Research Institute, Department of Physics and Astronomy, University of Delaware, Newark, DE 19716, USA\\
$^{4}$ Faculty of Engineering, Hokkai-Gakuen University, Toyohira-ku, Sapporo 062-8605, Japan}

\begin{document}

\pagerange{\pageref{firstpage}--\pageref{lastpage}} \pubyear{2011}

\maketitle

\label{firstpage}

\begin{abstract}
We present a three-dimensional (3-D) dynamical model for the broad [\altion{Fe}{iii}] emission observed in $\eta$ Carinae using the \emph{Hubble Space Telescope}/Space Telescope Imaging Spectrograph (\hst/STIS). This model is based on full 3-D Smoothed Particle Hydrodynamics (SPH) simulations of \ec's binary colliding winds. Radiative transfer codes are used to generate synthetic spectro-images of [\altion{Fe}{iii}] emission line structures at various observed orbital phases and STIS slit position angles (PAs). Through a parameter study that varies the orbital inclination $i$, the PA $\theta$ that the orbital plane projection of the line-of-sight makes with the apastron side of the semi-major axis, and the PA on the sky of the orbital axis, we are able, for the first time, to tightly constrain the absolute 3-D orientation of the binary orbit. To simultaneously reproduce the blue-shifted emission arcs observed at orbital phase 0.976, STIS slit $\mathrm{PA} = +38^{\circ}$, and the temporal variations in emission seen at negative slit PAs, the binary needs to have an $i \approx 130^{\circ}$ to $145^{\circ}$, $\theta \approx -15^{\circ}$ to $+30^{\circ}$, and an orbital axis projected on the sky at a $\mathrm{PA} \approx 302^{\circ}$ to $327^{\circ}$ east of north. This represents a system with an orbital axis that is closely aligned with the inferred polar axis of the Homunculus nebula, in 3-D. The companion star, \ecb, thus orbits clockwise on the sky and is on the observer's side of the system at apastron. This orientation has important implications for theories for the formation of the Homunculus and helps lay the groundwork for orbital modeling to determine the stellar masses.
\end{abstract}


\section{Introduction}

$\eta$ Carinae is the most luminous, evolved stellar object that can be closely studied \citep{davidson97}. Its immense luminosity \citep[$L \gtrsim 5 \times 10^{6} \ \Lsun$,][]{cox95} and relative proximity \citep[$D = 2.3 \pm 0.1 \ \mathrm{kpc}$,][]{smith06b} make it possible to test and constrain specific theoretical models of extremely massive stars ($\gtrsim 60 \ \Msun$) using high-quality data \citep{hillier01, hillier06, smithowocki06}. \ec\ is thus one of the most intensely observed stellar systems in the Galaxy, having been the focus of numerous ground- and space-based observing campaigns at multiple wavelengths from radio \citep{duncan03, white05} to gamma-rays \citep{tavani09, abdo10}.

Unfortunately, the dusty Homunculus nebula that formed during \ec's ``Great Eruption'' in the 1840s enshrouds the system, complicating direct observations of the central stellar source \citep{smith09}. Nevertheless, ground- and space-based, multi-wavelength observations obtained over the past two decades strongly indicate that \ec\ is a highly eccentric ($e\sim0.9$) colliding wind binary (CWB) with a 5.54-year orbital period \citep[][hereafter C11]{damineli96, feast01, duncan03, whitelock04, smith04, corcoran05, verner05, vangenderen06, damineli08a, damineli08b, eduardo10, corcoran11}. The consensus view is that the primary star, \eca, is a Luminous Blue Variable (LBV) \citep{hillier92, davidson97, hillier01, hillier06}. Radiative transfer modeling of \emph{Hubble Space Telescope}/Space Telescope Imaging Spectrograph (\hst/STIS) spatially-resolved spectroscopic observations suggests that \eca\ has a current mass $\gtrsim 90 \ M_{\odot}$, and a current-day stellar wind with a mass-loss rate of $\sim 10^{-3} M_{\odot} \ \mathrm{yr}^{-1}$ and terminal speed of $\sim 500 - 600 \ \mathrm{km \ s}^{-1}$ \citep[][hereafter H01, H06]{hillier01, hillier06}. The companion star, \ecb, continues to evade direct detection since \eca\ dwarfs its emission at most wavelengths (Damineli et al. 2000; H01; H06; Smith et al. 2004; C11). As a result, \ecb's stellar parameters, evolutionary state, orbit, and influence on the evolution of \eca\ are poorly known.

Currently, the best constraints on the stellar properties of \ecb\ come from photoionization modeling of the ``Weigelt blobs'', dense, slow-moving ejecta in the vicinity of the binary system \citep{weigelt86}. \citet{verner05} initially characterized \ecb\ as a mid-O to WN supergiant. More recent work by \citet{mehner10} tightly constrains the effective temperature of \ecb\ ($T_{\mathrm{eff}} \sim 37,000 - 43,000 \ \mathrm{K}$), but not its luminosity ($\log L/\Lsun \sim 5 - 6$), resulting in a larger range of allowed stellar parameters.

Constraints on the wind parameters of \ecb\ come from extended X-ray monitoring by the \emph{Rossi X-ray Timing Explorer} (\emph{RXTE}), \emph{Chandra}, \emph{XMM}, and \emph{Suzaku} satellites \citep{ishibashi99, corcoran05, hamaguchi07, corcoran10}. The periodic nature, minimum around periastron, and hardness (up to 10 keV) of \ec's \emph{RXTE} light curve are all characteristics of a highly eccentric CWB, the variable X-ray emission arising in a wind-wind collision (WWC) zone formed between the two stars \citep[][hereafter PC02, O08, P09, P11, and C10, respectively]{pittard02, okazaki08, parkin09, parkin10, corcoran10}. The hardness of the X-rays requires that \ecb\ have a high wind terminal speed of $\sim 3000 \ \mathrm{km} \ \mathrm{s}^{-1}$, and detailed modeling of the momentum balance between the two shock fronts suggests a mass-loss rate of $\sim 10^{-5} M_{\odot} \ \mathrm{yr}^{-1}$ (PC02; O08; P09; P11).

Proper numerical modeling of \ec's WWC remains a challenge, mainly because it requires a full three-dimensional (3-D) treatment since orbital motion, especially during periastron, can be important or even dominant, affecting the shape and dynamics of the WWC region (O08; P09; P11). Most hydrodynamical simulations of \ec\ have been two-dimensional (2-D), neglecting the effects of orbital motion for simplicity \citep[][PC02; Henley 2005]{pittard98, pittard00}. Until very recently (O08; P11), fully 3-D simulations were computationally impractical.

When it comes to reproducing observational diagnostics of \ec\ from hydrodynamical simulations of its colliding winds, the focus has been almost exclusively on X-rays. One drawback of this is that models for \ec\ at other wavelengths have been mostly phenomenological, offering only qualitative explanations for the vast array of complicated observations. This makes it difficult to test and refine models, or quantitatively constrain the physical parameters of the system. An excellent example is the lack of consensus regarding \ec's orbital orientation. The majority favor an orbit in which \ecb\ is behind \eca\ during periastron (Damineli 1996; PC02; Corcoran 2005; Hamaguchi et al. 2007; Nielsen et al. 2007a; Damineli et al. 2008b; Henley et al. 2008; O08; P09; P11; Groh et al. 2010b; Richardson et al. 2010). But others place \ecb\ on the near side of \eca\ at periastron \citep{falceta05, abraham05b, abrahamfalceta07, kashi07, kashi08, falceta09}. Some claim that the orientation is not well established at all \citep{mehner11}.

A more fundamental drawback of focusing purely on X-rays or other spatially-unresolved data is that any derived orbit is ambiguous with respect to its \emph{absolute} orientation on the sky; \emph{any orbital orientation derived solely from fitting the X-ray data can be rotated on the sky about the observer's line-of-sight and still match the observations}. A degeneracy also exists in the orbital inclination, with models that assume $i \approx 45^{\circ}$ and $i \approx 135^{\circ}$ both capable of fitting the observed \emph{RXTE} light curve. However, high-resolution spatial information is not enough; moderate-resolution spectral data is needed to determine the velocity structure of the emitting gas, and thus fully constrain the orbital orientation. As a result, using X-ray data alone, it is impossible to determine the 3-D alignment or misalignment of the orbital axis with the Homunculus polar axis, or the direction of the orbit. It is commonly \emph{assumed} that the orbital axis is aligned with the Homunculus polar axis, but to date, neither modeling nor observations have unambiguously demonstrated that this is the case. Knowing the orbit's orientation is key to constraining theories for the cause of the Great Eruption and formation of the Homunculus \citep[i.e. single star versus binary interaction/merger scenarios,][]{iben99, dwark02, smith03a, soker07, smith09, smith11, smithfrew10}. Furthermore, a precise set of orbital parameters would lay the groundwork for orbital modeling to determine the stellar masses.

In a recent attempt to characterize \ec's interacting winds, \citet[][hereafter G09]{gull09} presented an analysis of \emph{HST}/STIS observations taken between 1998 and 2004, identifying spatially-extended (up to $0.8''$), velocity-resolved forbidden emission lines from low- and high-ionization\footnote{Low- and high-ionization refer here to atomic species with ionization potentials (IPs) below and above the IP of hydrogen, 13.6 eV.} species. \emph{HST}/STIS imaging spectroscopy is ideal for moving beyond X-ray signatures as it uniquely provides the spatial ($\sim 0.1''$) \emph{and} spectral ($R \sim 8,000$) resolution necessary to separate the spectra of \ec's central source and nearby circumstellar ejecta from those of the Homunculus and other surrounding material. Moreover, this high-ionization forbidden emission is phase-locked, which strongly suggests it is regulated by the orbital motion of \ecb. Observed spectro-images are also highly dependent on the PA\footnote{Position Angle, measured in degrees from north to east.} of the \emph{HST}/STIS slit, implying that they contain valuable geometrical information. This opens up the possibility of obtaining \ecb's orbital, stellar, and wind parameters through proper modeling of the extended forbidden line emission.

This paper presents a detailed 3-D dynamical model for the broad components of the high-ionization forbidden line emission observed in \ec\ using the \emph{HST}/STIS. Our model is based on the results of full 3-D Smoothed Particle Hydrodynamics (SPH) simulations of \ec's colliding winds ($\S \ 3.2$ and $4.1$). Radiative transfer calculations performed with a modified version of the SPLASH code \citep{price07} are used with IDL routines to generate synthetic spectro-images of [\altion{Fe}{iii}] emission line structures at various orbital phases and STIS slit PAs ($\S \ 3.4$). Through a parameter study that varies the orbital inclination, the angle that the line-of-sight makes with the apastron side of the semi-major axis, and the PA on the sky of the projected orbital axis, we are able to tightly constrain, for the first time, the \emph{absolute} (3-D) orientation and direction of \ec's orbit ($\S \ 4$), showing in particular that the orbital axis is closely aligned in 3-D with the inferred polar axis of the Homunculus nebula. A discussion of the results and this derived orientation are in $\S \ 5$. $\S \ 6$ summarizes our conclusions and outlines the direction of future work. We begin ($\S \ 2$) with a brief summary of the observations used in this paper.

\section{The Observations}

\begin{figure*}
\includegraphics[width=17.5cm]{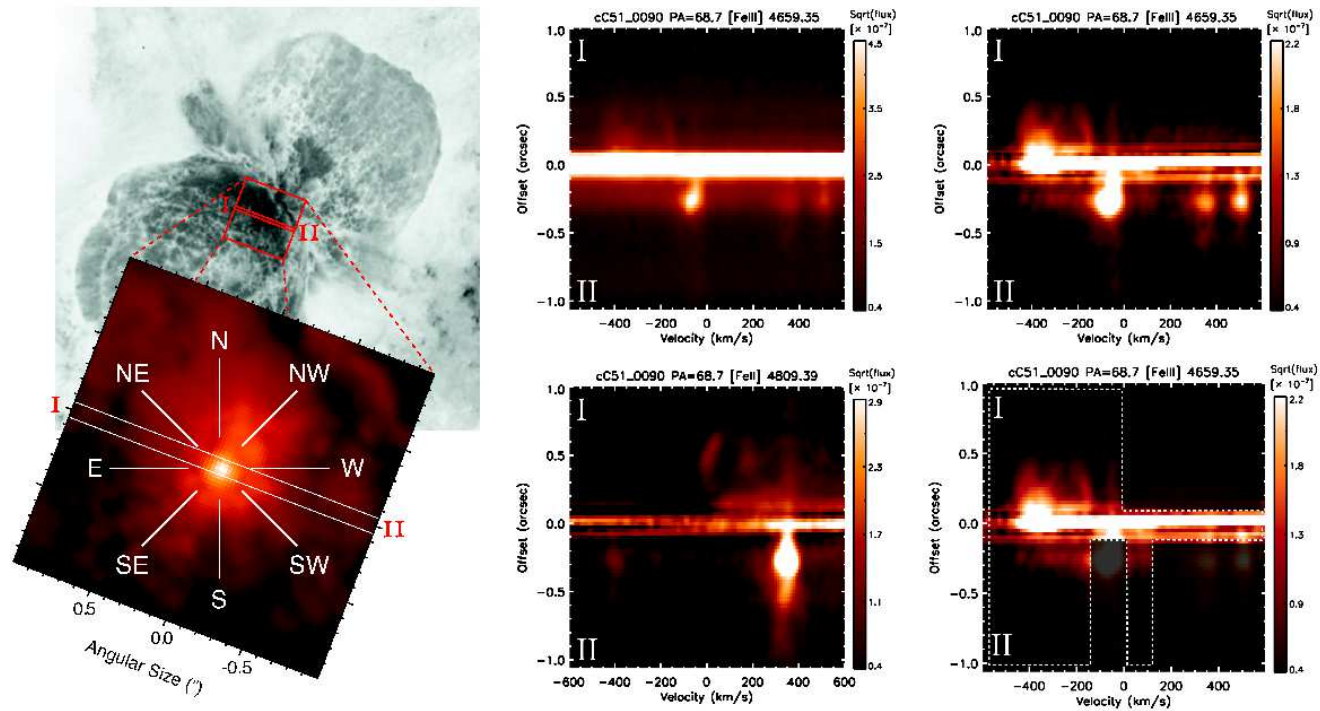}
\caption{Examples of the \emph{HST}/STIS spatially-resolved [\altion{Fe}{iii}] spectra. Top Left: An \emph{HST}/Advanced Camera for Surveys (ACS) High Resolution Camera (HRC) image of the Homunculus and \ec\ with a $20''$ field of view. Bottom Left: Enlarged central $2''$ field of view with the $52'' \times 0.1''$ slit positioned at PA = +69$^{\circ}$ and directional labels (NW = northwest, etc.) included for reference in the text. Spectro-images in the middle and right-hand columns were recorded from the central $2'' \times 0.1''$ portion of the slit, as drawn between labels I and II. Top Center: Spatially-resolved line profile of [\altion{Fe}{iii}] $\lambda$4659 in original form, recorded at $\mathrm{PA} = +69^{\circ}$ in 2002 July ($\phi = 0.82$). Top Right: The same spectro-image with the continuum subtracted on a spatial row-by-row basis. The velocity scale, referenced to the vacuum rest wavelength of the line, is $\pm 600 \ \mathrm{km} \ \mathrm{s}^{-1}$, and the color level is proportional to the square root of the intensity. Bottom Center: Spectro-image of [\altion{Fe}{ii}] $\lambda$4815, shifted by $400 \ \mathrm{km} \ \mathrm{s}^{-1}$, used as a template to determine the spatial and spectral ranges where the [\altion{Fe}{iii}] image is contaminated by [\altion{Fe}{ii}] $\lambda$4666 emission. Bottom Right: Same spectro-image as the top right panel, but with masks that remove the narrow line contamination and the contamination to the red from nearby [\altion{Fe}{ii}] $\lambda$4666 (see text).}
\label{fig1}
\end{figure*}

This paper uses the same \hst/STIS data described in G09 (see their Table 1). Extracted portions of observations recorded from March 1998 to March 2004 with the STIS CCD ($0.0507'' \ \mathrm{pixel}^{-1}$ scale) with medium dispersion gratings ($R \sim 8,000$), in combination with the $52'' \times 0.1''$ slit from 1,640 to 10,100 {\AA}, were used to sample the spectrum of the central core of \ec. Phases of the observations, $\phi$, are relative to the \emph{RXTE} X-ray minimum at 1997.9604 \citep{corcoran05}: $\mathrm{JD}_{obs} = \mathrm{JD} \ 2450799.792 + 2024.0 \times \phi$. Solar panel orientation requirements constrained the STIS slit PAs. The reduced STIS CCD spectra available through the STScI archives (\url{http://archive. stsci.edu/prepds/etacar/}) were used. Compass directions (NW = north-west, etc.) describe the spatial extent of the emission. All wavelengths are in vacuum and velocities are heliocentric. For details, see G09 and references therein.

Figure \ref{fig1} shows examples of resolved, broad emission from [\altion{Fe}{iii}] $\lambda 4659$. In the top center panel, faint spatially- and velocity-resolved emission can be seen against bright nebular emission and dust-scattered stellar radiation. Strong, narrow line emission from the SW of \ec\ is also seen. The bright continuum close to the wavelength of the wind line of interest is subtracted on a spatial row-by-row basis in order to isolate the fainter broad forbidden line emission (top right panel of Figure \ref{fig1}). All spectro-images in this paper have been processed identically using a portion of the spectrum with no bright narrow- or broad-line contamination. Row-by-row subtraction across the stellar position is less successful due to insufficient correction in the data reduction for small tilts of the spectrum on the CCD (G09).

The bottom right panel of Figure \ref{fig1} contains the same spectro-image of [\altion{Fe}{iii}] $\lambda 4659$ as the top right panel, but with masks that remove narrow line contamination from the Weigelt blobs, and contamination to the red due to the nearby [\altion{Fe}{ii}] $\lambda$4666 wind line. The masks are intended to remove ``distracting'' emission features not relevant to the work in this paper, displaying only the observed broad structures that are the focus of the modeling (outlined in white). To determine the spatial and spectral ranges where [\altion{Fe}{ii}] $\lambda$4666 contaminates the [\altion{Fe}{iii}] $\lambda 4659$ image, we use as a template the uncontaminated spectro-image of the bright [\altion{Fe}{ii}] $\lambda$4815 line (bottom center panel of Figure \ref{fig1}). We also compare the [\altion{Fe}{iii}] $\lambda 4659$ images to those of the [\altion{Fe}{iii}] $\lambda 4702$ and [\altion{N}{ii}] $\lambda 5756$ lines, which form in nearly identical conditions as they have very similar critical densities and IPs (Table 3 of G09). A detailed discussion of the observations and masking procedure for all spectro-images modeled in this paper is in Appendix A (available in the online version of this paper - see Supporting Information).

\subsection{The Key Observational Constraints Modeled}

Three key features observed in spectro-images of the high-ionization forbidden line emission can be used to constrain the 3-D orientation of \ec's orbit. Each provides important clues about the nature and orbital variation of \ec's interacting winds. We briefly summarize these below.

\subsubsection{Constraint 1: Emission Arcs at Slit PA = $+38^{\circ}$, $\phi = 0.976$}

\begin{figure}
\begin{center}
\includegraphics[width=8.5cm]{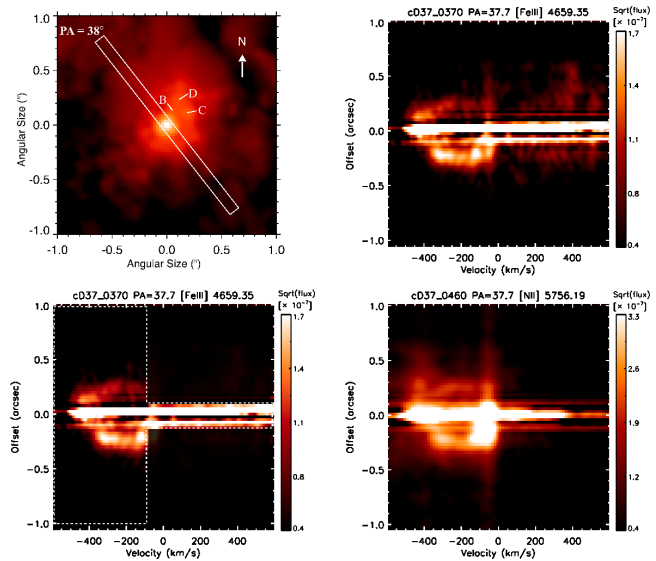}
\end{center}
\caption{Spectro-images recorded 2003 May 5 ($\phi = 0.976$). Top Left Panel: \emph{HST}/ACS HRC image of the central $2''$ of \ec\ with the $2''$ long portion of the STIS slit at PA = +38$^{\circ}$ overlaid. The locations of Weigelt blobs B, C, and D are indicated, as is the direction of north. Top Right Panel: Continuum-subtracted spectro-image of [\altion{Fe}{iii}] $\lambda$4659. Emission appears as a pair of completely blue-shifted arcs. Color is proportional to the square root of the intensity and the velocity scale is $\pm 600 \ \mathrm{km} \ \mathrm{s}^{-1}$. Weak emission to the red is entirely due to contamination by [\altion{Fe}{ii}] $\lambda$4666, while weak, narrow emission is centered at $\sim -40 \ \mathrm{km} \ \mathrm{s}^{-1}$. Bottom Left Panel: Same spectro-image, but with masks that remove the line contamination. Bottom Right Panel: Continuum-subtracted spectro-image of [\altion{N}{ii}] $\lambda$5756 included to demonstrate that at this phase and STIS PA, the high-ionization forbidden lines show no evidence of broad, spatially-extended, red-shifted emission.}
\label{fig2}
\end{figure}

Figure \ref{fig2} displays observed spectro-images of [\altion{Fe}{iii}] $\lambda$4659, recorded at slit PA = +38$^{\circ}$ on 2003 May 5 ($\phi = 0.976$), showing spatially-extended (up to $\sim 0.35''$) emission in the form of very distinct, nearly complete arcs that are entirely blue-shifted, up to $\sim -475 \ \mathrm{km} \ \mathrm{s}^{-1}$. Several weaker [\altion{Fe}{iii}] lines nearby in the spectrum, while blended, show no evidence of a spatially-extended ($> 0.1''$) red component, and no other high-ionization forbidden lines show an extended red component at this position (G09). To illustrate this, the bottom right panel of Figure \ref{fig2} contains a continuum-subtracted spectro-image of [\altion{N}{ii}] $\lambda$5756, taken at the same orbital phase and slit PA. The IP ($14.5 \ \mathrm{eV}$) and critical density ($3 \times 10^{7} \ \mathrm{cm}^{-3}$) of [\altion{N}{ii}] $\lambda$5756 are very similar to that of [\altion{Fe}{iii}] $\lambda$4659 ($16.2 \ \mathrm{eV}$ and $\sim 10^{7} \ \mathrm{cm}^{-3}$), meaning that their spatially-extended emission forms in nearly identical regions and via the same physical mechanism. The [\altion{N}{ii}] $\lambda$5756 line is strong and located in a spectral region with very few other lines. Unlike [\altion{Fe}{iii}] $\lambda$4659, there is no contamination to the red of [\altion{N}{ii}] $\lambda$5756, making it ideal for demonstrating that there is no spatially-extended, red-shifted, broad high-ionization forbidden line emission.

Of particular importance are the asymmetric shape and intensity of these emission arcs. Both arcs extend to nearly the same spatial distance from the central core, but the upper arc is noticeably dimmer and stretches farther to the blue, to $\sim -475 \ \mathrm{km} \ \mathrm{s}^{-1}$. The lower arc is brighter, but only extends to $\sim -400 \ \mathrm{km} \ \mathrm{s}^{-1}$. Every high-ionization forbidden line observed at this orbital phase and slit PA exhibits these asymmetries in intensity and velocity (figures $6 - 9$ of G09), indicating that they are independent of the IP and critical density of the line and intrinsic to the shape and distribution of the photoionized extended wind material in which the lines form. As such, the shape and asymmetry of the blue-shifted emission arcs may be used to help constrain the orbital orientation.

\subsubsection{Constraint 2: Variations with Orbital Phase at Constant Slit PA = $-28^{\circ}$}

Let us focus next on shifts of the high-ionization forbidden emission with orbital phase. Figure 11 of G09 shows sets of six spectra of [\altion{Fe}{iii}] $\lambda$4659 and $\lambda$4702 recorded at select phases at slit $\mathrm{PA} = -28^{\circ}$. During the two minima at $\phi = 0.045$ and 1.040 (rows 1 and 5 of figure 11 of G09), the broad [\altion{Fe}{iii}] emission is absent. By $\phi = 0.213$ (row 2) and $\phi = 1.122$ (row 6, $\sim 8$ months after the X-ray minimum), it strongly reappears. Images at $\phi = 0.407$ are similar to those at $\phi = 0.213$ and 1.122. By $\phi = 0.952$, emission above $\sim 150 \ \mathrm{km} \ \mathrm{s}^{-1}$ disappears. This phase dependence in the [\altion{Fe}{iii}] emission, especially the disappearance during periastron, strongly indicates that it is tied to the orbital motion of \ecb.

Another important observational feature is the presence of spatially-extended \emph{red-shifted} emission to the NW, in addition to extended blue-shifted emission to the SE, at phases far from periastron. This is in contrast to the entirely blue-shifted arcs seen for slit PA = $+38^{\circ}$, indicating that the STIS slit at PA = $-28^{\circ}$ is sampling different emitting portions of \ec's extended wind structures, even at similar orbital phases.

\subsubsection{Constraint 3: Doppler Shift Correlations with Orbital Phase and Slit PA}

The final set of observations focuses on changes with slit PA leading up to periastron. Figures 12 and 13 of G09 illustrate the behavior of the [\altion{Fe}{iii}] emission for various slit PAs at select phases. The key point to take away from this particular data set is that away from periastron, the spatially-extended [\altion{Fe}{iii}] emission is almost entirely blue-shifted for positive slit PAs = $+22^{\circ}$ to $+70^{\circ}$, but is partially red-shifted for negative PAs, most notably PA = $-82^{\circ}$.

\section{A 3-D Dynamical Model for the Broad High-Ionization Forbidden Line Emission}

\subsection{Why Model [\altion{Fe}{iii}] $\lambda$4659 Emission?}

Table 3 of G09 lists the various forbidden emission lines observed in \ec. However, there are several important reasons to focus particular attention on [\altion{Fe}{iii}] $\lambda$4659 emission.

First, the physical mechanism for the formation of forbidden lines is well understood (Appendix B, online version). Moreover, emission from forbidden lines is optically thin. Thus, forbidden line radiation can escape from a nebula much more easily than radiation from an optically thick resonance line, which is emitted and reabsorbed many times \citep{hartman03}. This is of great value when studying an object so enshrouded by circumstellar material since one does not have to model complicated radiative transfer effects.

[\altion{Fe}{iii}] is considered rather than a lower ionization line of [\altion{Fe}{ii}] because \emph{the bulk of the} [\altion{Fe}{iii}] \emph{emission arises in regions that are directly photoionized by \ecb} (Verner et al. 2005; G09; Mehner et al. 2010). No intrinsic [\altion{Fe}{iii}] emission is expected from \eca. This is based on detailed theoretical models by H01 and H06, which show that in \eca's envelope, Fe$^{2+}$ only exists in regions where the electron density is two-to-four orders of magnitude higher than the critical density of the [\altion{Fe}{iii}] $\lambda$4659 line. Since Fe$^{0}$ needs $16.2 \ \mathrm{eV}$ radiation (or collisions) to reach the Fe$^{2+}$ state, most of the [\altion{Fe}{iii}] emission arises in areas directly photoionized by \ecb.

In comparison, Fe$^{0}$ only needs 7.9 eV radiation or collisions to ionize to Fe$^{+}$, which can form in the wind of \eca\ (without \ecb's influence) and/or near the wind-wind interaction regions in areas excited by mid-UV radiation filtered by \eca's dense wind. Collisions and photoexcitation to upper Fe$^{+}$ energy levels can also populate many metastable levels. As a result, [\altion{Fe}{ii}] emission is far more complicated and originates from lower excitation, lower density regions on much larger spatial scales (H01; H06; G09; Mehner et al. 2010). Because of this complexity, the modeling of [\altion{Fe}{ii}] emission is deferred to future work.

The IP of 16.2 eV required for Fe$^{2+}$ also means that the [\altion{Fe}{iii}] emission forms in regions where hydrogen is ionized, but helium is still neutral. In contrast, both Ar$^{2+}$ and Ne$^{2+}$ have IPs greater than $24.6$ eV. Therefore, [\altion{Ar}{iii}] and [\altion{Ne}{iii}] emission arise in areas where helium is singly ionized. Modeling [\altion{Fe}{iii}] is thus more straightforward as one does not have to worry about the two different types of ionization structure possible (one due to H and one due to He), which depend on the spectrum of ionizing radiation and the abundance of helium \citep{osterbrock89}.

[\altion{Fe}{iii}] $\lambda$4659 is chosen over the similar emission line of [\altion{N}{ii}] $\lambda$5756 in order to avoid potential complications due to intrinsic [\altion{N}{ii}] emission in the extended wind of \eca. [\altion{N}{ii}] emission can form in a broad zone between 100 and 1000$R_{*}$ ($\sim 30 - 300$ AU) of \eca\ and is sensitive to its wind temperature and mass-loss rate (H01). By focusing on [\altion{Fe}{iii}], one avoids having to include any intrinsic forbidden line emission from \eca.

The modeling in this paper focuses solely on the \emph{broad} [\altion{Fe}{iii}] emission features that are thought to arise in the dense, moderate velocity ($\sim 100 - 600 \ \mathrm{km \ s}^{-1}$) extended primary wind and WWC regions. The much narrower ($\lesssim 50 \ \mathrm{km \ s}^{-1}$) emission features that form in the Weigelt blobs and other dense, slow-moving equatorial ejecta from \ec's smaller eruption in the 1890s \citep{weigelt86, davidson95, davidsonetal97, ishibashi03, smith04, gull09, mehner10} are not modeled. Any possible effects of local dust formation are neglected for simplicity as not enough information is available to realistically include them at this time \citep{williams08, mehner10}.

As all observed high-ionization forbidden lines show the same basic spatial and temporal features in their broad emission, with any differences in size or location attributable to differences in the IP and critical density of the specific line of interest, focusing solely on [\altion{Fe}{iii}] $\lambda$4659 should not significantly bias the overall conclusions, which should extend to the other high-ionization forbidden lines as well. Values of the transition probability ($A_{21} = 0.44 \ \mathrm{s}^{-1}$), statistical weights ($g_{1} = g_{2} = 9$), and line transition frequency ($\nu_{21} = 6.435 \times 10^{14} \ \mathrm{s}^{-1}$) for [\altion{Fe}{iii}] $\lambda$4659 all come from \citet{nahar96}. Values for the collision strengths ($\Omega_{12}$) come from \citet{zhang96}. A solar abundance of iron is also used, consistent with the works of H01; \citet{verner05}; and H06.

\subsection{The 3-D SPH Code and General Problem Setup}

The numerical simulations in this paper were performed with the same 3-D SPH code used in O08. The stellar winds are modeled by an ensemble of gas particles that are continuously ejected with a given outward velocity at a radius just outside each star. For simplicity, both winds are taken to be adiabatic, with the same initial temperature ($35,000$ K) at the stellar surfaces and coasting without any net external forces, assuming that gravitational forces are effectively canceled by radiative driving terms (O08). The gas has negligible self-gravity and adiabatic cooling is included.

In a standard $xyz$ Cartesian coordinate system, the binary orbit is set in the $xy$ plane, with the origin at the system center-of-mass and the orbital major axis along the $x$-axis. The two stars orbit counter-clockwise when looking down on the orbital plane along the $+z$ axis. Simulations are started with the stars at apastron and run for multiple consecutive orbits. By convention, $t = 0$ ($\phi = 0$) is defined to be at periastron passage.

To match the $\sim \pm 0.7''$ scale of the STIS observations at the adopted distance of 2.3 kpc to the \ec\ system \citep{smith06b}, the outer simulation boundary is set at $r = \pm 105a$ from the system center-of-mass, where $a$ is the length of the orbital semi-major axis ($a = 15.4$ AU). Particles crossing this boundary are removed from the simulation. We note that the SPH formalism is ideally suited for such large-scale 3-D simulations, as compared to grid-based hydrodynamics codes \citep{price04, monaghan05}.

Table \ref{tab4.1} summarizes the stellar, wind, and orbital parameters used in the modeling, which are consistent with those derived from the observations (PC02; H01; H06; C11), with the exception of the wind temperature of \eca. The effect of the wind temperature on the dynamics of the high-velocity wind collision is negligible (O08). Note that the simulation adopts the higher mass-loss rate for \eca\ derived by \citet{davidson95}, \citet{cox95}, H01, and H06, rather than the factor of four lower mass-loss rate assumed when analyzing X-ray signatures (PC02; O08; P09; P11; C11). Effects of the adopted value of the primary mass-loss rate on the forbidden line emission will be the subject of a future paper.

The publicly available software SPLASH \citep{price07} is used to visualize the 3-D SPH code output. SPLASH differs from other tools because it is designed to visualize SPH data using SPH algorithms. There are a number of benefits to using SPLASH, and the reader is referred to \citet{price07} and the SPLASH userguide for these and discussions on the interpolation algorithms.

\begin{table}
\caption{Stellar, Wind, and Orbital Parameters of the 3-D SPH Simulation}
\label{tab4.1}
\begin{center}
\begin{tabular}{l c c}\hline
  Parameter & $\eta_{\mathrm{A}}$ & $\eta_{\mathrm{B}}$ \\ \hline
  Mass ($M_{\odot}$) & 90 & 30 \\
  Radius ($R_{\odot}$) & 90 & 30 \\
  Mass-Loss Rate ($M_{\odot}$ yr$^{-1}$) & $10^{-3}$ & $10^{-5}$ \\
  Wind Terminal Velocity (km s$^{-1}$) & 500 & 3000 \\
  Orbital Period (days) & \multicolumn{2}{c}{2024} \\
  Orbital Eccentricity $e$ & \multicolumn{2}{c}{0.9} \\
  Semi-major Axis Length $a$ (AU) & \multicolumn{2}{c}{15.4}  \\\hline
\end{tabular}
\end{center}
\end{table}

\subsection{Generation of Synthetic Slit Spectro-Images}

The 3-D SPH simulation provides the time-dependent, 3-D density and temperature structure of \ec's interacting winds on spatial scales comparable to those of the \emph{HST}/STIS observations. This forms the basis of our 3-D dynamical model. Unfortunately, it is currently not possible to perform full 3-D simulations of \ec\ in which the radiative transfer is properly coupled to the hydrodynamics \citep{paardekooper10}. Instead, the radiative transfer calculations here are performed as post-processing on the 3-D SPH simulation output. This should not strongly affect the results so long as the material that is photoionized by \ecb\ responds nearly instantaneously to its UV flux. In other words, as long as the timescale for the recombination of $e + \mathrm{Fe}^{2+} \rightarrow \mathrm{Fe}^{+}$ is very small relative to the orbital timescale, the calculations should be valid.

The recombination timescale is $\tau_{\mathrm{rec}} = 1/(\alpha_{\mathrm{rec}} n_{e})$ seconds, where $\alpha_{\mathrm{rec}}(T)$ is the recombination rate coefficient (in units of $\mathrm{cm}^{3} \ \mathrm{s}^{-1}$) at temperature $T$ and $n_{e}$ is the electron number density. For $T \sim 10^{4}$ K, $\alpha_{\mathrm{rec}}(T) \approx 4 \times 10^{-12} \ \mathrm{cm}^{3} \ \mathrm{s}^{-1}$ \citep{nahar97}. Since densities near the critical density of the line are the primary focus, $n_{e} \sim 10^{7} \ \mathrm{cm}^{-3}$; this gives $\tau_{\mathrm{rec}} \approx 7 \ \mathrm{hours}$, which is much smaller than the orbital timescale, even during periastron passage, which takes approximately one month. Any possible time-delay effects should not strongly affect the results as the light-travel-time to material at distances of $\sim 0.5''$ in \ec\ is only about one week.

\subsubsection{Summary of the Basic Procedure}

Synthetic slit spectro-images are generated using a combination of Interactive Data Language (IDL) routines and radiative transfer calculations performed with a modified version of the SPLASH \citep{price07} code. Here, we outline the basic procedure, with the specifics discussed in the following subsections.

Nearly all of the required calculations are performed within SPLASH, which reads in the 3-D SPH code output for a specific orbital phase, rotates the data to a desired orbital orientation on the sky, and computes the line-of-sight velocity for all of the material. Next is the computation of the ionization volume created by \ecb\ where Fe$^{+}$ is photoionized to Fe$^{2+}$. It is assumed that any material within this volume is in collisional ionization equilibrium, and that no material with $T > 250,000$ K emits. All other material in the photoionization volume has its forbidden line emissivity calculated using Equation (\ref{5.23}), derived below. Material located outside the photoionization volume does not emit.

The intensity $I$ is computed by performing a line-of-sight integration of the emissivity at each pixel, resulting in an image of the spatial distribution of the [\altion{Fe}{iii}] intensity projected on the sky. This is done for each velocity $v_{bin}$ used along the dispersion axis of the synthetic slit in the velocity range of interest, resulting in a set of intensity images. The individual [\altion{Fe}{iii}] $\lambda$4659 intensity images are then combined and convolved with the \emph{HST}/STIS response using IDL routines in order to create a synthetic position versus velocity spectro-image for comparison to the observations.

\subsubsection{Defining the Binary Orientation Relative to the Observer}

\begin{figure}
\begin{center}
\includegraphics[width=7.7cm]{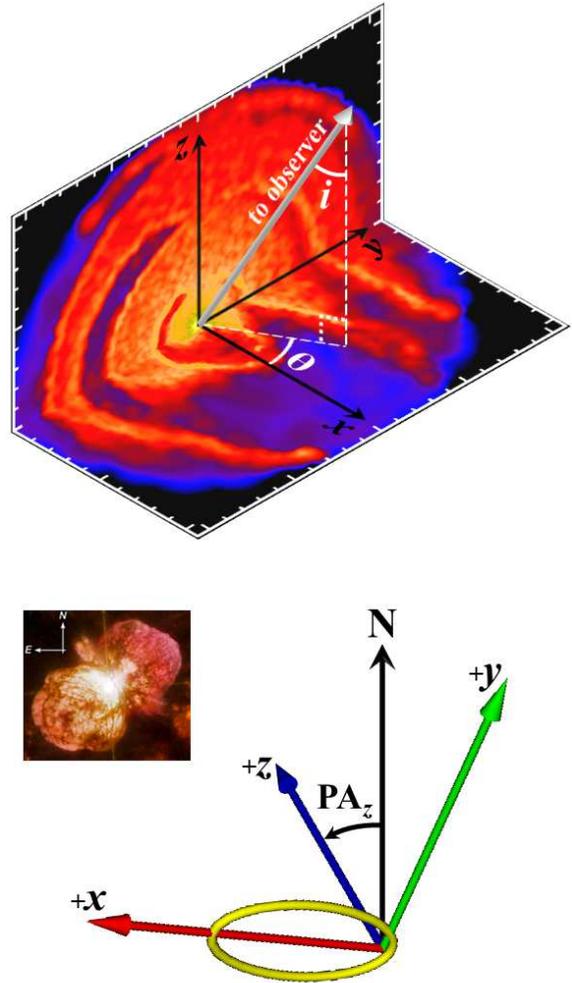}
\end{center}
\caption{Diagrams illustrating the observer's position. Top: Schematic defining the inclination angle $i$ that the observer's line-of-sight makes with the $+z$ orbital axis, and the equatorial projection angle $\theta$ of the line-of-sight relative to the apastron side of the semi-major axis $x$. The background orthogonal planes show slices of density from the 3-D SPH simulation at apastron in the $xy$ orbital plane and the $yz$ plane perpendicular to the orbital plane and major axis. Bottom: Diagram defining the position angle on the sky, $\mathrm{PA}_{z}$, of the $+z$ orbital axis (blue), measured in degrees counter-clockwise of north (N). The binary orbit projected on the sky is shown in yellow, as are the semi-major (red) and semi-minor (green) axes. The small inset to the left is a \hst\ WFPC2 image of the Homunculus (Credit: NASA, ESA, and the Hubble SM4 ERO Team) included for reference.}
\label{orientation}
\end{figure}

With the two stars orbiting in the $xy$ plane, the 3-D orientation of the binary relative to the observer is defined by the orbital inclination $i$, the prograde direction angle $\theta$ that the orbital plane projection of the observer's line-of-sight makes with the apastron side of the semi-major axis\footnote{$\theta$ is the same as the angle $\phi$ defined in figure 3 of O08.}, and the position angle \emph{on the sky} of the $+z$ orbital axis, PA$_{z}$ (Figure \ref{orientation}).

Following the standard definition for binary systems, $i$ is the angle that the line-of-sight makes with the $+z$ axis. An $i = 0^{\circ}$ is a face-on orbit with the observer's line-of-sight along the $+z$ axis and the two stars orbiting counter-clockwise on the sky. An $i = 90^{\circ}$ places the line-of-sight in the orbital $xy$ plane, while $i = 180^{\circ}$ is a face-on orbit with the two stars orbiting clockwise on the sky.

For $i = 90^{\circ}$, a value of $\theta = 0^{\circ}$ has the observer looking along the $+x$ axis (along the semi-major axis on the apastron side of the system), while a $\theta = 90^{\circ}$ has the observer looking along the $+y$ axis. In the conventional notation of binary orbits, the `argument of periapsis' $\omega = 270^{\circ} - \theta$.

PA$_{z}$ defines the position angle on the sky of the $+z$ orbital axis and is measured in degrees counter-clockwise of N. A $\mathrm{PA}_{z} = 312^{\circ}$ aligns the projected orbital axis with the Homunculus polar axis \citep{davidson01, smith06b, smith09} and has $+z$ pointing NW on the sky. A $\mathrm{PA}_{z} = 42^{\circ}$ places the orbital axis perpendicular to the Homunculus polar axis with $+z$ pointing NE. A $\mathrm{PA}_{z} = 132^{\circ}$ $(222^{\circ}$) is also aligned with (perpendicular to) the Homunculus polar axis and has $+z$ pointing SE (SW) on the sky.

\subsubsection{Volume of Material Photoionized by \ecb}

Based on the models of \eca's envelope by H01 and H06, it is assumed that Fe$^{+}$ is initially the dominant ionization state of iron in primary wind material located $r > 50 \ \mathrm{AU}$ from \eca, and that hydrogen and helium are initially neutral in the primary wind for distances of $r > 155 \ \mathrm{AU}$ and $> 3.7 \ \mathrm{AU}$, respectively. The calculation of the volume of material photoionized from Fe$^{+}$ to Fe$^{2+}$ by \ecb\ follows that presented in \citet{nussbaumer87} for symbiotic star systems. For cases where hydrogen or helium alone determine the ionization structure, the boundary between neutral hydrogen (H$^{0}$) and ionized hydrogen (H$^{+}$) can be expressed analytically. As \eca's extended wind consists mostly of hydrogen and helium (figure 9 of H01), and because the [\altion{Fe}{iii}] emission arises in regions where hydrogen is ionized but helium is not, such a calculation should provide a reasonable approximation for the size and shape of the photoionization boundary between Fe$^{+}$ and Fe$^{2+}$. This, however, is an upper limit for the photoionization volume since the IP of 16.2 eV required for Fe$^{2+}$ is slightly larger than the 13.6 eV needed to ionize H$^{0}$.

In our calculation, the primary star is separated by a distance $p$ from a hot companion star that emits spherically symmetrically $L_{\mathrm{H}}$ photons s$^{-1}$ that are capable of ionizing H$^{0}$. The primary has a standard spherically symmetric mass-loss rate of $\dot{M}_{1} = 4\pi r^{2}\mu m_{\mathrm{H}} n(r)v_{\infty}$, where $n(r)$ is the hydrogen number density, $\mu$ is the mean molecular weight, $m_{\mathrm{H}}$ is the mass of a hydrogen atom, and $v_{\infty}$ is the terminal speed of the primary wind.

In a small angle $\Delta \theta$ around the direction $\theta$, the equilibrium condition for recombination-ionization balance is

\begin{equation}
L_{\mathrm{H}} \frac{\Delta \theta}{4\pi} = \Delta \theta \int_{0}^{s_{\theta}} n(s) n_{e}(s) \alpha_{B}(\mathrm{H},T_{e}) \ s^{2} \ ds , \label{5.24}
\end{equation}\\
where $s$ measures the distance from the ionizing secondary star in the direction $\theta$, $s_{\theta}$ is the boundary between H$^{0}$ and H$^{+}$, $\alpha_{B}$ is the total hydrogenic recombination coefficient in case B at electron temperature $T_{e}$, and $n_{e}$ is the electron number density \citep{nussbaumer87}. The electron density can be written as

\begin{equation}
n_{e}(r) = (1 + a(\mathrm{He})) n(r) \ , \label{5.25}
\end{equation}\
where
\begin{equation}
n(r) = \frac{\dot{M}_{1}}{4\pi r^{2} \mu m_{\mathrm{H}} v_{\infty}} \ , \label{5.26}
\end{equation}\\
and $a(\mathrm{He})$ is the abundance by number of He relative to H. Expressing both $s$ and $r$ in units of $p$, defining $u \equiv s/p$, and making the further approximation that $\alpha_{B}$ is constant \citep{nussbaumer87}, Equation (\ref{5.24}) can be written as

\begin{equation}
X^{\mathrm{H}^{+}} = f(u,\theta) \ , \label{5.27}
\end{equation}\
where
\begin{equation}
X^{\mathrm{H}^{+}} = \frac{4\pi\mu^{2}m_{\mathrm{H}}^{2}}{\alpha_{B}(1 + a(\mathrm{He}))} p L_{\mathrm{H}} \left(\frac{v_{\infty}}{\dot{M}_{1}} \right)^{2} \ , \label{5.28}
\end{equation}\\
and
\begin{equation}
f(u,\theta) = \int_{0}^{u} \frac{x^{2}}{(x^{2} - 2x\cos\theta + 1)^{2}} \ dx \ . \label{5.29}
\end{equation}\

In the modified SPLASH routine, Equation (\ref{5.27}) is solved for a specified $L_{\mathrm{H}}$. However, since Equation (\ref{5.27}) does not account for the effects of the WWC, we assume that mass is conserved and that all primary wind material normally within the cavity created by \ecb\ is compressed into the walls of the WWC region, with the ionization state of the WWC zone walls before being photoionized the same as that of the primary wind (i.e. iron is in the Fe$^{+}$ state).

The stellar separation is computed directly from the 3-D SPH simulation. We use the same value for the abundance by number of He to H as H01 and H06, $a(\mathrm{He}) = 0.2$. A constant value of $\alpha_{B} = 2.56 \times 10^{-13} \ \mathrm{cm}^{3} \ \mathrm{s}^{-1}$ at $T_{e} = 10,000$ K is also used \citep{osterbrock89}. Based on \citet{mehner10} and \citet{verner05}, the value of $L_{\mathrm{H}}$ is chosen to be that of an O5 giant with $T_{eff} \approx 40,000 \ \mathrm{K}$, which according to \citet{martins05} has a hydrogen ionizing flux of $10^{49.48}$ photons s$^{-1}$. Any shielding effects due to the presence of the Weigelt blobs or other dense, slow-moving equatorial ejecta are not presently included.

The third column of Figure \ref{fig5} illustrates the time variability of the photoionization region created by \ecb\ for slices in the $xy$ orbital plane. The region is largest around apastron, when \ecb\ is farthest from \eca's dense wind. As \ecb\ moves closer to \eca, the photoionization zone becomes smaller and more wedge-shaped\footnote{The outer edge looks circular only because this marks the edge of the spherical computational domain of the SPH simulation.} due to the gradual embedding of \ecb\ in \eca's wind. During periastron, \ecb\ becomes completely enshrouded in \eca's thick wind, preventing material at large distances ($\gtrsim 10a$) from being photoionized. As \ecb\ emerges after periastron, the photoionization region is restored and grows as \ecb\ moves back toward apastron. This plunging into, and withdrawal from, \eca's wind by \ecb\ leads to the illumination of distant material in very specific directions as a function of phase.

\subsubsection{The Emissivity Equation}

For a simple two-level atom, the volume emissivity $j$ (erg cm$^{-3}$ s$^{-1}$ sr$^{-1}$) of a forbidden line is

\begin{equation}
j = \frac{1}{4\pi} h \nu_{21} N_{2} A_{21} \ , \label{5.15}
\end{equation}\\
where $N_{2}$ is the number density of atoms in the excited upper level and $\nu_{21}$ is the frequency of the line transition \citep{dopita03}. Using the total number density of element of interest $E$ in ionization state $i$, $n_{i,E} \approx N_{1} + N_{2}$ \citep{ignace06}, together with Equation (B5), one finds

\begin{equation}
N_{2} = \left(\frac{n_{e}q_{12}}{n_{e}q_{21} + A_{21}} \right) (n_{i,E} - N_{2}) \ . \label{5.16}
\end{equation}\\
Solving for $N_{2}$ and using Equation (B4) for $q_{21}/q_{12}$ gives

\begin{equation}
N_{2} = n_{i,E} \left\{1 + \frac{g_{1}}{g_{2}}\exp\left(\frac{h \nu_{21}}{kT}\right) \left[1 + \frac{n_{c}}{n_{e}} \right] \right\}^{-1} \ , \label{5.17}
\end{equation}\\
where $n_{c}$, defined in Equation (B6), is the critical density of the line. By defining $Q_{i,E} \equiv n_{i,E}/n_{E}$ as the fraction of element $E$ in ionization state $i$, $A_{E} \equiv n_{E}/n_{N}$ as the abundance of element $E$ relative to all nucleons $n_{N}$, and $\gamma_{e} \equiv n_{N}/n_{e}$ as the ratio of nucleons to electrons \citep{ignace06}, $n_{i,E}$ becomes

\begin{equation}
n_{i,E} = Q_{i,E} A_{E} \gamma_{e} n_{e} \ . \label{5.18}
\end{equation}\
Therefore, using Equations (\ref{5.17}) and (\ref{5.18}),

\begin{eqnarray}
j & = & \frac{h \nu_{21} A_{21} Q_{i,E} A_{E}\gamma_{e} n_{e}}{4\pi} \nonumber\\ \nonumber\\
& & \times \left\{1 + \frac{g_{1}}{g_{2}} \exp \left(\frac{h \nu_{21}}{kT}\right) \left[1 + \frac{n_{c}}{n_{e}}\right] \right\}^{-1} \ . \label{5.19}
\end{eqnarray}\

Based on Equation (\ref{5.19}), emission from a specific forbidden line is concentrated only in regions that (1) contain the right ionization state of the element of interest (in our case, Fe$^{2+}$), (2) are near the critical density of the line (for [\altion{Fe}{iii}], $\sim 10^{7} \ \mathrm{cm}^{-3}$), and (3) are near the right temperature (for [\altion{Fe}{iii}], $T \approx 32,000 \ \mathrm{K}$). This important point is crucial to understanding the forbidden line emission observed in \ec.

To compute the emissivity using SPLASH, the line profiles are assumed to have a Gaussian thermal broadening, so that Equation (\ref{5.19}) is weighted by an exponential with the form

\begin{eqnarray}
j & = & \frac{h \nu_{21} A_{21} Q_{i,E} A_{E}\gamma_{e} n_{e}}{4\pi} \exp\left[-\left(v_{bin} - \frac{v_{los}}{v_{th}} \right)^{2} \right] \nonumber\\ \nonumber\\
& & \times \left\{1 + \exp\left(\frac{h \nu_{21}}{kT}\right) \left[\frac{g_{1}}{g_{2}} + \frac{A_{21}g_{1}T^{1/2}}{\beta \Omega_{12} n_{e}} \right] \right\}^{-1}\ , \label{5.22}
\end{eqnarray}\\
where Equation (B8) has been used for $n_{c}$, $v_{los}$ is the line-of-sight velocity of material in the 3-D SPH simulation for a specified orientation of the binary orbit relative to the observer, $v_{th}$ is a thermal velocity dispersion (taken to be 25 km s$^{-1}$), and $v_{bin}$ is the bin size used along the dispersion axis of the synthetic slit in the wavelength range of interest, expressed in thermal velocity units (25 km s$^{-1}$).

To relate the physical density in the SPH simulation to the electron number density, we use $n_{e} = \rho_{sph}/(\mu_{e}m_{\mathrm{H}})$, with $\mu_{e}$ the mean molecular weight per free electron. As the focus is on forbidden lines of trace metals, the value of $n_{e}$ should not be impacted by the ionization balance of the metals \citep{ignace06}. Moreover, since far-UV radiation (or high-energy collisions) is necessary for the formation of the high-ionization forbidden lines, wherever [\altion{Fe}{iii}] emission occurs, hydrogen should also be ionized. For such regions dominated by H$^{+}$, we thus take $\mu_{e} = \gamma_{e} = 1$. The volume emissivity then takes the final general form,

\begin{eqnarray}
j & = & \frac{h \nu_{21} A_{21} Q_{i,E} A_{E} \rho_{sph}}{4\pi m_{H}} \exp\left[-\left(v_{bin} - \frac{v_{los}}{v_{th}} \right)^{2} \right] \nonumber\\ \nonumber\\
& & \times \left\{1 + \exp\left(\frac{h\nu_{21}}{kT}\right) \left[\frac{g_{1}}{g_{2}} + \frac{A_{21}g_{1}T^{1/2}m_{\mathrm{H}}}{\beta \Omega_{12} \rho_{sph}} \right] \right\}^{-1} . \label{5.23}
\end{eqnarray}\\

The appropriate value of $Q_{i,E}$ is found assuming collisional ionization equilibrium, with the fraction of Fe$^{2+}$ as a function of $T$ based on the ion fraction data from \citet{bryans06}.
These data show that at $T \sim 250,000$ K, the fractional abundance of Fe$^{2+}$ is $10^{-5.453}$, with the remaining Fe being collisionally ionized to higher states. The amount of Fe$^{2+}$ decreases even more at higher temperatures. The model therefore assumes zero [\altion{Fe}{iii}] emission from material with $T > 250,000$ K.

\subsubsection{The Intensity and Synthetic Slit Spectro-Images}

Since the [\altion{Fe}{iii}] emission is optically thin, the intensity \emph{I} is simply the integral of the volume emissivity $j$ along the line-of-sight,
\begin{equation}
I = \int j \ dl \label{5.31}
\end{equation}\\
\citep{mihalas78}. The intensity is computed in SPLASH by performing a line-of-sight integration of $j$ through the entire 3-D simulation at each pixel, resulting in an image of the intensity in the [\altion{Fe}{iii}] $\lambda$4659 line centered at a particular velocity (wavelength), as an observer would see it projected on the sky. This is done for multiple velocities ($v_{bin}$) in $25 \ \mathrm{km} \ \mathrm{s}^{-1}$ intervals for line-of-sight velocities from $-600 \ \mathrm{km} \ \mathrm{s}^{-1}$ to $+600 \ \mathrm{km} \ \mathrm{s}^{-1}$.

The resulting series of [\altion{Fe}{iii}] intensity images are combined using IDL routines to create a synthetic position versus velocity spectro-image. The IDL code reads in the SPLASH images and rotates them to a specified PA on the sky corresponding to a desired PA of the \emph{HST}/STIS slit, assuming that the orbital axis is either aligned with the Homunculus polar axis at a $\mathrm{PA}_{z} = 312^{\circ}$, or rotated relative to the Homunculus axis by some specified angle. Each image is cropped to match the $0.1''$ width of the STIS slit, with the slit assumed centered exactly on the \ec\ central source. Since position information is only available along one direction, each image has its intensity values integrated along each row of pixels within the slit width. This produces a single `slice' of the [\altion{Fe}{iii}] $\lambda$4659 intensity along the slit, centered at a specific velocity. These slices are combined to create the synthetic spectro-image.

The resulting model spectro-images have spatial ($\sim 0.003''$) and spectral (25 km s$^{-1}$) resolutions that are better than the \emph{HST} observations. The synthetic spectro-images are therefore convolved with the response of \emph{HST}/STIS in order to match its spatial ($0.1''$) and spectral ($37.5 \ \mathrm{km} \ \mathrm{s}^{-1}$) resolutions. Point-spread functions (PSFs) for STIS generated using the Tiny Tim program \citep{krist99} are used for the convolution in the spatial direction, while a gaussian is used for the convolution in the spectral direction. Color in the model spectro-images scales as the square root of the intensity, with the colorbar ranging from 0 to $1/4$ of the maximum intensity, the same as used for displaying the observations.

\begin{figure}
\begin{center}
\includegraphics[width=8.5cm]{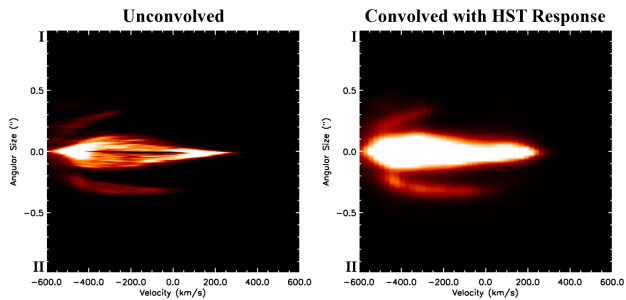}
\end{center}
\caption{Example synthetic spectro-images of [\altion{Fe}{iii}] $\lambda$4659 both unconvolved (left) and convolved with the response of \emph{HST}/STIS (right), assuming $i = 138^{\circ}$, $\theta = 0^{\circ}$, and $\mathrm{PA}_{z} = 312^{\circ}$. Roman numerals I and II indicate the top and bottom of the slit, respectively. Color is proportional to the square root of the intensity, and the velocity scale is $\pm 600 \ \mathrm{km} \ \mathrm{s}^{-1}$.}
\label{unconvolved}
\end{figure}

Figure \ref{unconvolved} shows that there is a noticeable difference between the unconvolved (left) and convolved (right) model spectro-images, especially in the central $\pm 0.15''$. Details in the unconvolved spectro-images that arise near the inner WWC zone are completely unresolved in the convolved images, resulting in a bright, central streak of emission extending from negative to positive velocities. This is consistent with the observations and indicates that \emph{HST}/STIS lacks the spatial resolution needed to resolve the details of \ec's inner WWC zone.

\section{Results from the 3-D Dynamical Model}

In the discussions below, it is assumed for simplicity that phase zero of the spectroscopic cycle (from the observations) coincides with phase zero of the orbital cycle (periastron passage). In a highly-eccentric binary system like \ec, the two values are not expected to be shifted by more than a few weeks \citep{groh10a}. Such a time shift would only cause a small change of $\sim 10^{\circ}$ in the derived best value of $\theta$, which will not affect the overall conclusions.

\subsection{Hydrodynamics of the Extended WWC Region}

The first two columns of Figures \ref{fig4} and \ref{fig5} show, respectively, the density and temperature in the $xy$ orbital plane from the 3-D SPH simulation. The binary system has undergone multiple orbits, as indicated by the dense arcs of wind material in the outer ($> 20a$) regions. On the $-x$ (periastron) side of the system are narrow cavities carved by \ecb\ in \eca's dense wind during each periastron passage. The cavities are most easily seen in the temperature plots as they contain warm ($\sim 10^{4} - 10^{5} \ \mathrm{K}$), low-density wind material from \ecb. Bordering these cavities are compressed, high-density shells of primary wind that form as a result of the WWC.

The insets of Figure \ref{fig5} illustrate the creation of one of these narrow wind cavities and its bordering dense shell of primary wind. Their spiral shape is due to the increased orbital speeds of the stars during periastron. This is in contrast to phases around apastron (rows b, c and d) when orbital speeds are much lower and the current WWC zone maintains a simple, axisymmetric conical shape (O08; P09; P11). The increasing orbital speeds when approaching periastron causes the postshock gas in the leading arm of the WWC zone to be heated to higher temperatures than the gas in the trailing arm (insets of rows d and e), an affect also found by P11.

Following periastron, \ecb\ moves back to the $+x$ (apastron) side of the system, its wind colliding with and heating the dense primary wind that flows unimpeded in the $+x,-y$ direction. As pointed out by P11, the arms of the WWC region shortly after this ($\phi \approx 0.1$) are so distorted by orbital motion that the leading arm collides with the trailing arm from before periastron, leading to additional heating of the postshock gas in the trailing arm (compare $\phi \approx 1.122$, Figure~\ref{fig5}, row f). The leading arm of the WWC zone (including the portion that collides with the trailing arm) helps to form a dense, compressed shell of primary wind that flows in the $+x$ and $-y$ directions after periastron passage. Because the wind of \ecb\ collides with a dense wall of postshock primary wind with high inertia, the primary wind controls the overall rate of expansion of the resulting spiral (P11).

The overall stability of the expanding shell of primary wind on the apastron side of the system depends on the shock thickness (Vishniac 1983; W\"{u}nsch et al. 2010; P11). Portions of the shell moving in the $-y$ direction appear to be the most stable due to the increased amount of primary wind that borders it in this direction. However, our simulations show that at $\phi \approx 0.3$, the upper portion of the shell expanding in the $+x,+y$ direction starts to fragment. Eventually, the wind of \ecb\ is able to plough through the shell, causing it to separate from the leading arm of the WWC zone ($\phi \approx 0.4$, row c of Figure \ref{fig5}). This produces a pair of dense `arcs' of primary wind on the apastron side of the system. The arc expanding in the $+x,-y$ direction is the remnant of the leading arm of the WWC region from the previous periastron passage. The arc in the $+x, +y$ direction is the remnant of the trailing arm of the WWC region from just \emph{prior} to the previous periastron passage. Multiple pairs of these arcs are visible in Figures \ref{fig4} and \ref{fig5}. They are also quite spatially extended; by the time the system is back at periastron, the arcs from the previous periastron are up to $70a$ from the central stars (in the $xy$ plane). As the arcs expand further, they gradually mix with the surrounding low-density wind material from \ecb.

The fragmentation of the shell and formation of the arcs is mainly due to the above-mentioned collision of the leading arm of the WWC zone with the trailing arm just after periastron. This collision produces instabilities at the interface between the two arms (P11), which, together with the shell's expansion, the pressure from \ecb's high-velocity wind, and the lack of bordering primary wind material for support, causes the shell to break apart in the $+x,+y$ direction $\sim 1.7$ years after periastron.

While the term `shell' is used to describe the layer of compressed primary wind that flows in the direction of apastron following periastron passage, this is not related to the `shell ejection' event discussed in the past \ec\ literature \citep{zanella84, davidson99, davidsonetal99, smith03a, davidson05}. The scenario proposed by \citet{zanella84} and advocated by \citet{davidson99} to explain \ec's spectroscopic events is a qualitative single-star model wherein some sort of thermal or surface instability is presumed to induce significant mass loss from \ec\ approximately every 5.54 years. A shell ejection as the result of a latitude-dependent disturbance in the wind of a single star has also been proposed \citep{smith03a, davidson05}. \citet{davidson05} suggested that the shell event may be triggered by the close approach of a secondary star, but this is still mainly a single-star scenario requiring some kind of surface instability in the primary.

In contrast to these ideas, the formation of the outflowing shell of primary wind seen in the 3-D SPH simulations is a natural, unavoidable consequence in any high-eccentricity, massive CWB in which the primary star has a significant mass-loss rate (significant both in terms of overall mass-loss rate and mass-loss rate relative to the companion star). It is not so much an `ejection' event as it is a chance for the normal primary wind to flow in the direction of apastron for a brief time ($\sim 3 - 5$ months) while \ecb\ performs periastron passage. Eventually, \ecb\ returns to the apastron side and its high-velocity wind collides with and compresses this primary wind material, forming a dense shell that continues to propagate in the direction of apastron until it breaks apart at its weakest point, resulting in a pair of dense `arcs', as described above.

\subsection{Physical Origin and Location of the Broad High-Ionization Forbidden Line Emission}

\begin{figure*}
\includegraphics[width=17.5cm]{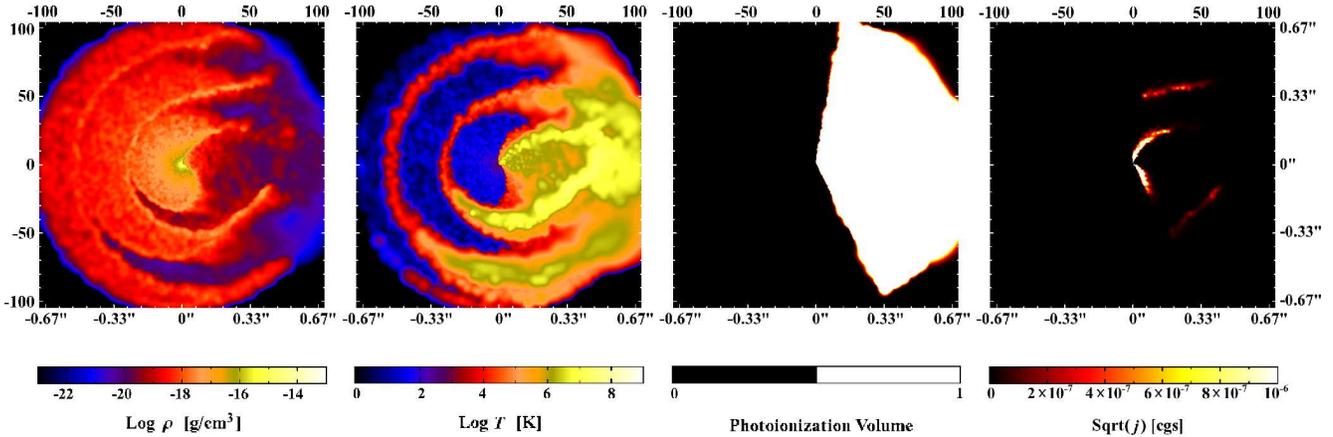}
\caption{Snapshots in the orbital $xy$ plane from the 3-D SPH simulation of \ec\ at $\phi = 0.976$ used to model the observed blue-shifted emission arcs of Figure~\protect\ref{fig2}. Color shows, from left to right, log density, log temperature, photoionization volume created by \ecb\ (white = ionized), and square root of the modeled emissivity of the [\altion{Fe}{iii}] $\lambda$4659 line. The color bar in the last panel has been adjusted to make faint emission more visible. The box size is $\pm 105a \approx \pm 1622 \ \mathrm{AU}$ $\approx \pm 0.7''$ ($D = 2.3$ kpc). Axis tick marks correspond to an increment of $10a \approx 154 \ \mathrm{AU} \approx 0.067''$. In the last panel, [\altion{Fe}{iii}] emission only originates from material within the photoionization volume that is near the critical density of the [\altion{Fe}{iii}] line ($\sim 10^{7} \ \mathrm{cm}^{-3}$) and at the appropriate temperature.}
\label{fig4}
\end{figure*}

Figure \ref{fig4} presents slices in the orbital plane from the 3-D SPH simulation at $\phi = 0.976$, used to model the blue-shifted emission arcs seen at STIS slit $\mathrm{PA} = +38^{\circ}$. The colors show log density, log temperature, photoionization zone created by \ecb, and square root of the modeled emissivity of the [\altion{Fe}{iii}] $\lambda$4659 line. The photoionization region is confined to the same side of the system as \ecb, the dense wind of \eca\ preventing the far ($-x$) side of the system from being photoionized. All of the [\altion{Fe}{iii}] emission is thus concentrated in two spatially-distinct regions on the apastron side of the system:
\begin{itemize}
  \item The strongest [\altion{Fe}{iii}] emission originates near the walls of the current WWC zone in the inner $\sim 30a \approx 0.2''$ of \ec.
  \item Faint, spatially-extended (out to $\sim 60a \approx 0.4''$) [\altion{Fe}{iii}] emission arises in the arcs of dense primary wind formed during the previous periastron passage.
\end{itemize}
The [\altion{Fe}{iii}] emission occurs in these two areas because they are the regions within the photoionization zone that are near the critical density of the [\altion{Fe}{iii}] line and at the appropriate temperature.

\begin{figure*}
\includegraphics[width=17.5cm]{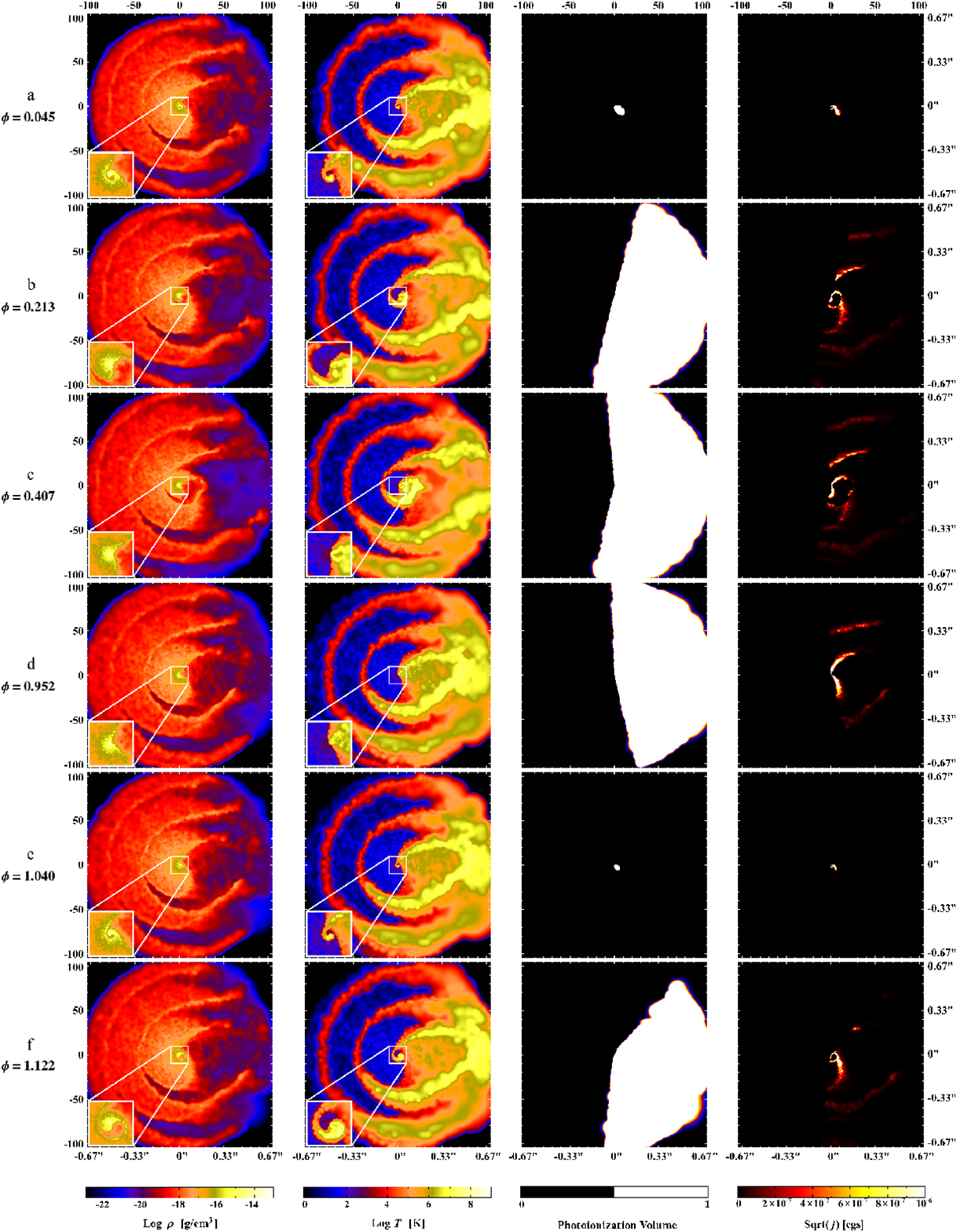}
\caption{Same as Figure \protect\ref{fig4}, but for $\phi = 0.045$, 0.213, 0.407, 0.952, 1.040, and 1.122 (rows, top to bottom), which correspond to phases at which \hst/STIS observations were accomplished (figure~11 of G09). Insets in the first two columns are a zoom of the inner $10a$, included to illustrate the complex dynamics of the `current' WWC region.}
\label{fig5}
\end{figure*}

Figure \ref{fig5} illustrates how the density, temperature, photoionization region, and [\altion{Fe}{iii}] $\lambda$4659 emissivity change with orbital phase. The six phases shown correspond to those observed in figure 11 of G09. During periastron passage ($\phi = 0.045$ and 1.040, rows a and e), \ecb\ becomes deeply embedded in the dense wind of \eca\ and the photoionization region shrinks considerably, down to only a few semi-major axes across. This leads to an effective `shutting-off' of the forbidden line emission at large distances ($\gtrsim 10a$). By $\phi = 0.213$ and 1.122 (rows b and f), \ecb\ has moved far enough from \eca\ to re-establish the large photoionization zone and spatially-extended forbidden line emission. The photoionization region is largest, and the [\altion{Fe}{iii}] most spatially extended, around apastron (row c, $\phi = 0.407$). By $\phi = 0.952$ (row d), \ecb\ starts to become embedded in the dense wind of \eca, the photoionization region in the $+x, -y$ quadrant decreasing slightly in size.

Far-UV radiation from \ecb\ leads to highly ionized regions that extend outward from its low density wind cavity in the direction of system apastron. During periastron passage, the disappearance of the extended high-ionization forbidden emission can be attributed to the wrapping of the dense primary wind around \ecb, which traps its far-UV radiation and prevents it from photoionizing the outer wind structures responsible for the observed emission. While \ecb\ is embedded in \eca's wind, the latter flows unimpeded in the direction of apastron, eventually forming dense arcs of material that expand outward as \ecb\ completes periastron passage and the inner regions of the WWC zone regain their near conical shape. These cooler, dense arcs of expanding primary wind also produce high-ionization forbidden line emission when photoionized by \ecb.

Observed temporal variations of the high-ionization forbidden lines can thus be linked to the orbital motion of \ecb\ in its highly eccentric orbit, which causes different portions of the WWC regions and extended arcs of primary wind from earlier cycles to be photoionized. As the two stars move closer to or farther from each other, orbital motion also leads to changes in the density and temperature of the inner WWC zone, which in turn modifies the size and shape of \ecb's photoionization volume, and thus the overall shape, location, and intensity of the high-ionization forbidden line emission, even at phases away from periastron (rows b, c, d, and f of Figure \ref{fig5}). Forbidden line emission from the compressed, inner WWC region increases in intensity until the critical density is reached or the temperature exceeds that at which the appropriate ions can exist. Forbidden emission from the extended arcs of primary wind decreases in intensity as the arcs gradually expand and mix with the surrounding low-density wind material from \ecb.

\subsection{Synthetic Slit Spectro-images and Constraining the Orbital Orientation}

\subsubsection{Constraint 1: Emission Arcs at Slit PA = $+38^{\circ}$, $\phi = 0.976$}

The blue-shifted emission arcs in Figure \ref{fig2} represent distinct, well-defined structures observed at a specific orbital phase and slit PA. As such, they provide a natural basis for modeling and can be used to constrain the orbital orientation. We have performed a parameter study in $i$, $\theta$, and $\mathrm{PA}_{z}$ with the goal of determining which set(s), if any, of orientation parameters result in synthetic spectro-images that closely match the observations in Figure \ref{fig2}. The value of $\theta$ was varied in $15^{\circ}$ increments for values of $0^{\circ} \leq \theta \leq 360^{\circ}$, with $i$ varied in $5^{\circ}$ increments over the range $0^{\circ} \leq i \leq 180^{\circ}$ for each $\theta$ value, and $\mathrm{PA}_{z}$ varied in $5^{\circ}$ increments over the range $0^{\circ} \leq \mathrm{PA}_{z} \leq 360^{\circ}$ for each pair of $i, \theta$ values.

\begin{figure*}
\includegraphics[width=17.5cm]{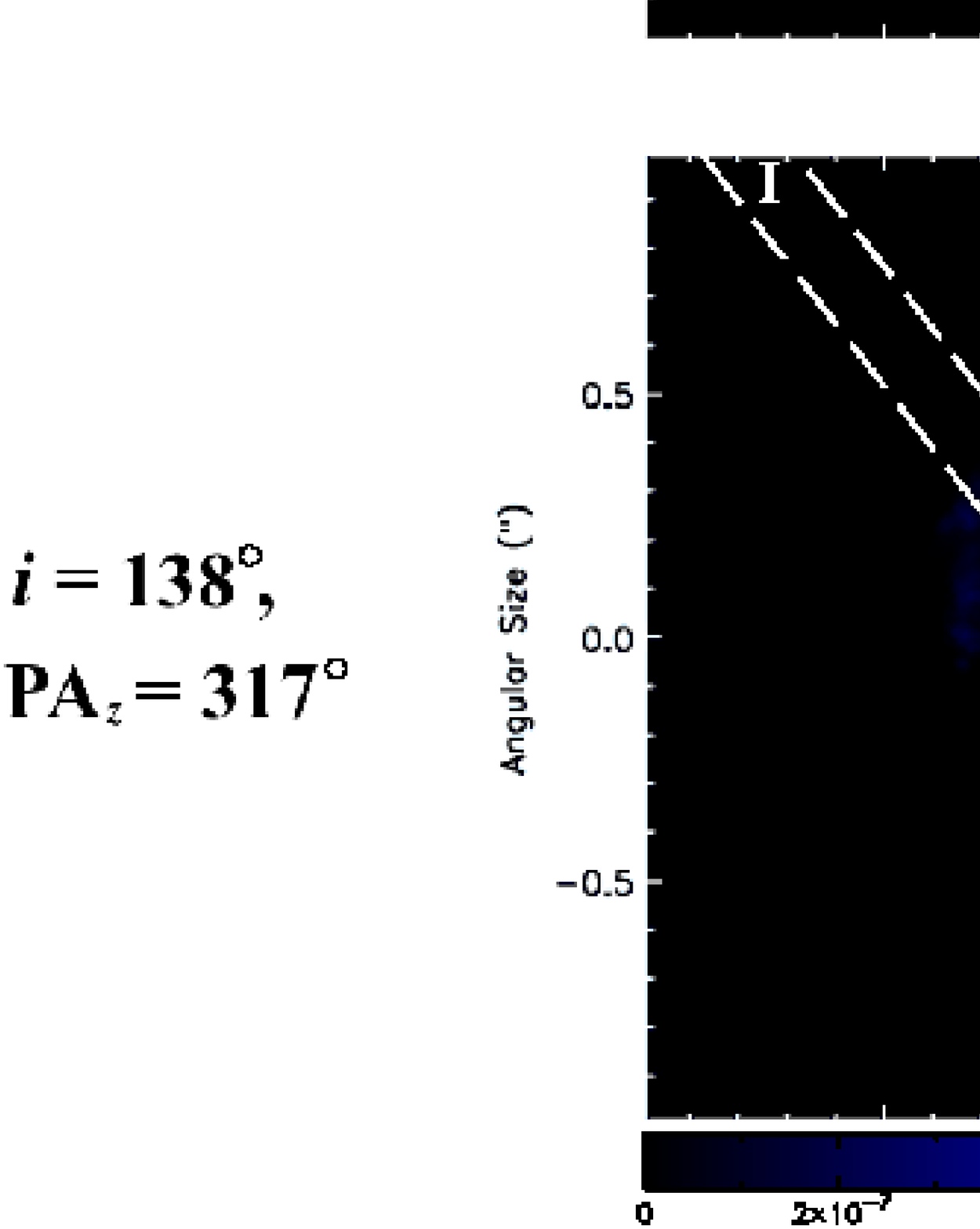}
\caption{2-D spatial distribution on the sky of the square root of the modeled intensity in the [\altion{Fe}{iii}] $\lambda$4659 line and synthetic spectro-images for orbital orientations $i = 42^{\circ}$, $\theta = 7^{\circ}$, $\mathrm{PA}_{z} = 302^{\circ}$ (top row) and $i = 138^{\circ}$, $\theta = 7^{\circ}$, $\mathrm{PA}_{z} = 317^{\circ}$ (bottom row), which lie near the centers of the two derived best-fit ranges of orientation parameters for matching the observations taken at $\phi = 0.976$, $\mathrm{PA} = 38^{\circ}$ in Figure \protect\ref{fig2}. Columns are, from left to right: 2-D distribution on the sky of the square root of the modeled intensity in the [\altion{Fe}{iii}] $\lambda$4659 line for blue-shifted material with $v_{los}$ between $-500 \ \mathrm{km} \ \mathrm{s}^{-1}$ and $-100 \ \mathrm{km} \ \mathrm{s}^{-1}$; same as first column, but for red-shifted material with $v_{los}$ between $+100 \ \mathrm{km} \ \mathrm{s}^{-1}$ and $+500 \ \mathrm{km} \ \mathrm{s}^{-1}$; model spectro-image convolved with the response of \emph{HST}/STIS; and observed spectro-image with mask. The projected $x$, $y$, and $z$ axes are shown for reference in the 2-D images of the intensity on the sky, as is the direction of north. The $0.1''$ wide STIS slit at $\mathrm{PA} = +38^{\circ}$ is also overlaid. Roman numerals I and II indicate the top and bottom of the slit, respectively. All following 2-D projections of the modeled [\altion{Fe}{iii}] intensity in this paper use this same labeling convention. The color scale in the spectro-images is proportional to the square root of the intensity and the velocity scale is $\pm 600 \ \mathrm{km} \ \mathrm{s}^{-1}$. Lengths are in arcseconds.}
\label{fig9}
\end{figure*}

\begin{figure*}
\includegraphics[width=17.7cm]{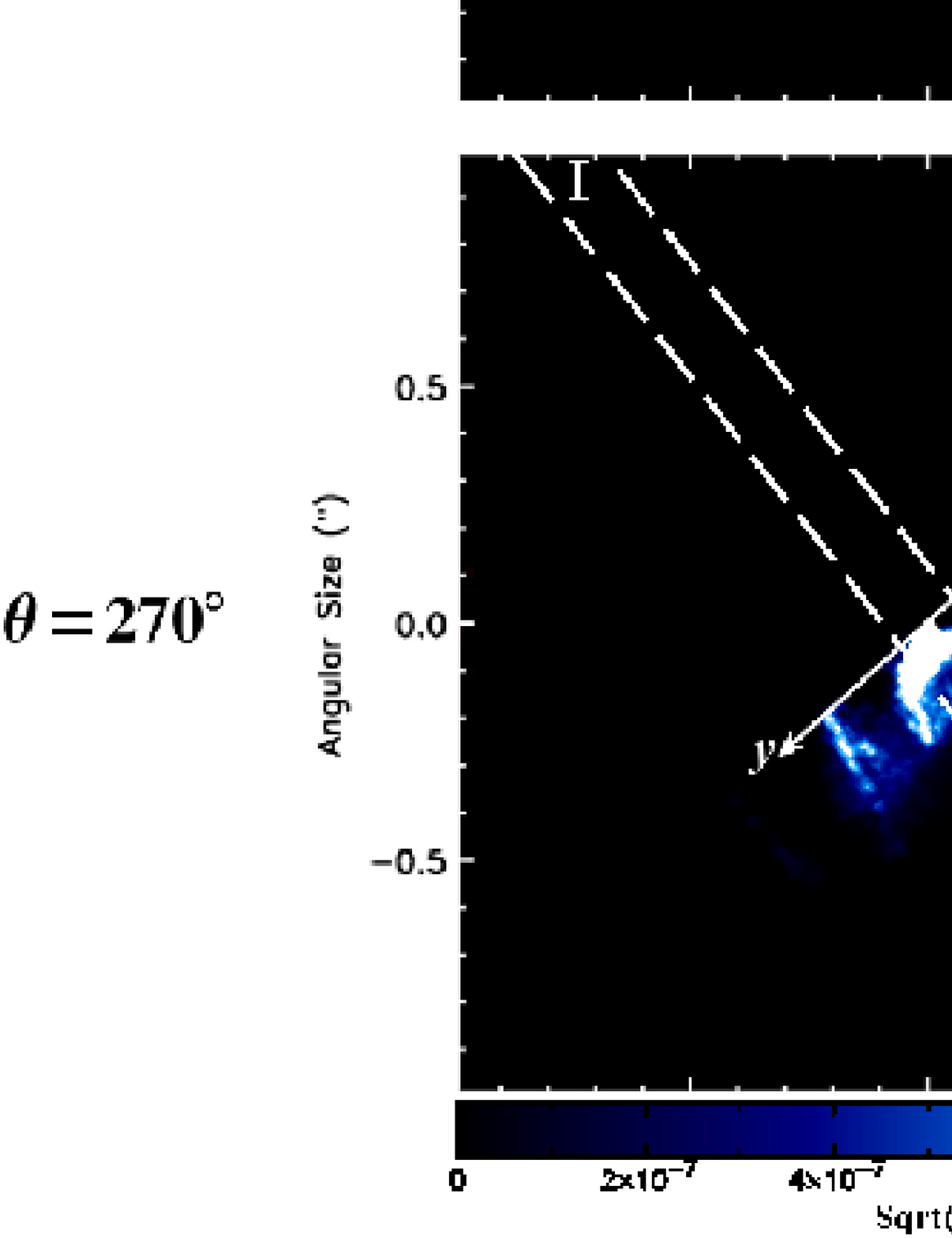}
\caption{Same as Figure~\ref{fig9}, but for values of  $\theta = 0^{\circ}, \ 90^{\circ}, \ 180^{\circ}, \ \mathrm{and} \ 270^{\circ}$ (rows, top to bottom), assuming $i = 138^{\circ}$ and $\mathrm{PA}_{z} = 312^{\circ}$.}
\label{fig6}
\end{figure*}

We find that \emph{only} synthetic spectro-images generated for $-15^{\circ} \leq \theta \leq +45^{\circ}$ are able to reasonably match the observations taken at $\phi = 0.976$, $\mathrm{PA} = 38^{\circ}$. However, an ambiguity exists in $i$ and $\mathrm{PA}_{z}$, with values of $30^{\circ} \leq i \leq 50^{\circ}$, $\mathrm{PA}_{z} = 272^{\circ}$ to $332^{\circ}$ (top row of Figure~\ref{fig9}) and $130^{\circ} \leq i \leq 150^{\circ}$, $\mathrm{PA}_{z} = 282^{\circ}$ to $342^{\circ}$ (bottom row of Figure~\ref{fig9}) both capable of producing entirely blue-shifted arcs that resemble those observed. Our best morphological fits are obtained using values of $i = 42^{\circ}$, $\theta = 7^{\circ}$, $\mathrm{PA}_{z} = 302^{\circ}$ or $i = 138^{\circ}$, $\theta = 7^{\circ}$, $\mathrm{PA}_{z} = 317^{\circ}$, which lie near the centers of these two derived ranges of best-fit orientation parameters.

The observed morphology and asymmetries in brightness and $v_{los}$ of the spatially-extended emission arcs determine the allowed ranges of the orbital orientation parameters. The parameters $i$ and $\mathrm{PA}_z$ primarily control the brightness asymmetry and spatial orientation of the blue- and red-shifted emission components on the sky. Figure~\ref{fig9} shows that the synthetic spectro-images at slit $\mathrm{PA} = 38^{\circ}$ for $i = 42^{\circ}$ and $138^{\circ}$ (at nearly identical $\theta$ and $\mathrm{PA}_z$) are remarkably similar. Yet, the two orbital orientations are drastically different, resulting in unique distributions on the sky of the blue- and red-shifted emission. The red component is the most notable, which extends to the SE for $i = 42^{\circ}$, but to the NW for $i = 138^{\circ}$. The asymmetry in brightness between the two blue-shifted arcs is also controlled by $i$.  Only values of $30^{\circ} \leq i \leq 50^{\circ}$ or $130^{\circ} \leq i \leq 150^{\circ}$ produce a brightness asymmetry, with other $i$ resulting in arcs that are approximately equal in brightness.

The value of $\mathrm{PA}_z$ determines the final orientation of the projected $z$ axis, and thus of the individual emission components. Because the lower blue-shifted arc in the observations is brighter than the upper arc, the extended blue emission projected on the sky must be brighter in directions to the SW (where the lower half of the slit is located, position II in Figure~\ref{fig9}), and dimmer in directions to the NE (position I). This limits $\mathrm{PA}_{z}$ to values between $272^{\circ}$ and $342^{\circ}$.

More importantly, $i$ and  $\mathrm{PA}_z$ determine the 3-D orientation of the orbital axis. If $i = 42^{\circ}$ and $\mathrm{PA}_{z} = 272^{\circ}$ to $342^{\circ}$, the orbital axis is inclined toward the observer and \emph{is not aligned} with the Homunculus polar axis. The two stars would also orbit counter-clockwise on the sky. However, if $i = 138^{\circ}$, then, because $\mathrm{PA}_{z} \approx 312^{\circ}$, the orbital axis \emph{is} closely aligned \emph{in 3-D} with the Homunculus polar axis, and the stars orbit clockwise on the sky.

Figure \ref{fig6} shows that the value of $\theta$ (for fixed $i = 138^{\circ}$ and $\mathrm{PA}_{z} = 312^{\circ}$) determines the $v_{los}$ of the material exhibiting [\altion{Fe}{iii}] emission. According to our 3-D model, the spatially-extended [\altion{Fe}{iii}] emission originates in the expanding arcs of dense primary wind formed during the previous periastron passage. When $\theta \approx 0^{\circ}$ (top row of Figure~\ref{fig6}) this emission is mostly blue-shifted because the emitting material within \ecb's photoionization zone is moving mostly toward the observer. The model spectro-image consists of spatially-extended, blue-shifted arcs because the slit at $\mathrm{PA} = 38^{\circ}$ primarily samples the blue component, which stretches from NE to SW on the sky. In contrast, the extended red component is \emph{not} sampled; only a small amount of red-shifted emission in the very central $\pm 0.1''$ core falls within the slit. Moreover, only values of $-15^{\circ} \leq \theta \leq +45^{\circ}$ produce significant amounts of spatially-extended ($\sim 0.7''$ in total length) blue-shifted emission with the correct observed asymmetry in $v_{los}$, wherein the dimmer upper arc extends $\sim 75 \ \mathrm{km} \ \mathrm{s}^{-1}$ more to the blue than the brighter lower arc.

Figure \ref{fig9} demonstrates that our 3-D dynamical model and derived best-fit orbital orientation(s) reproduce all of the key features in the observations taken at $\phi = 0.976$, $\mathrm{PA} = 38^{\circ}$. Both synthetic and observed spectro-images contain concentrated, velocity-extended emission in the central $\pm 0.1''$ core. The completely blue-shifted emission arcs extend to roughly the same spatial distances ($\sim \pm 0.35''$). The asymmetry in brightness between the upper and lower arcs is matched as well. The asymmetry in $v_{los}$ is also reproduced and of the same magnitude. However, the arcs in the model images stretch a bit farther to the blue, to $\sim -550 \ \mathrm{km} \ \mathrm{s}^{-1}$ (upper arc) and $-475 \ \mathrm{km} \ \mathrm{s}^{-1}$ (lower arc), versus $\sim -475 \ \mathrm{km} \ \mathrm{s}^{-1}$ and $-400 \ \mathrm{km} \ \mathrm{s}^{-1}$ in the observations. The synthetic images thus show emission at velocities slightly above the value used in the SPH simulation for \eca's wind terminal speed. This is a minor discrepancy though, likely due to the extended emitting material receiving an extra `push' from the fast wind of \ecb\ (see subsection 4.3.4).

\begin{figure*}
\includegraphics[width=17.5cm]{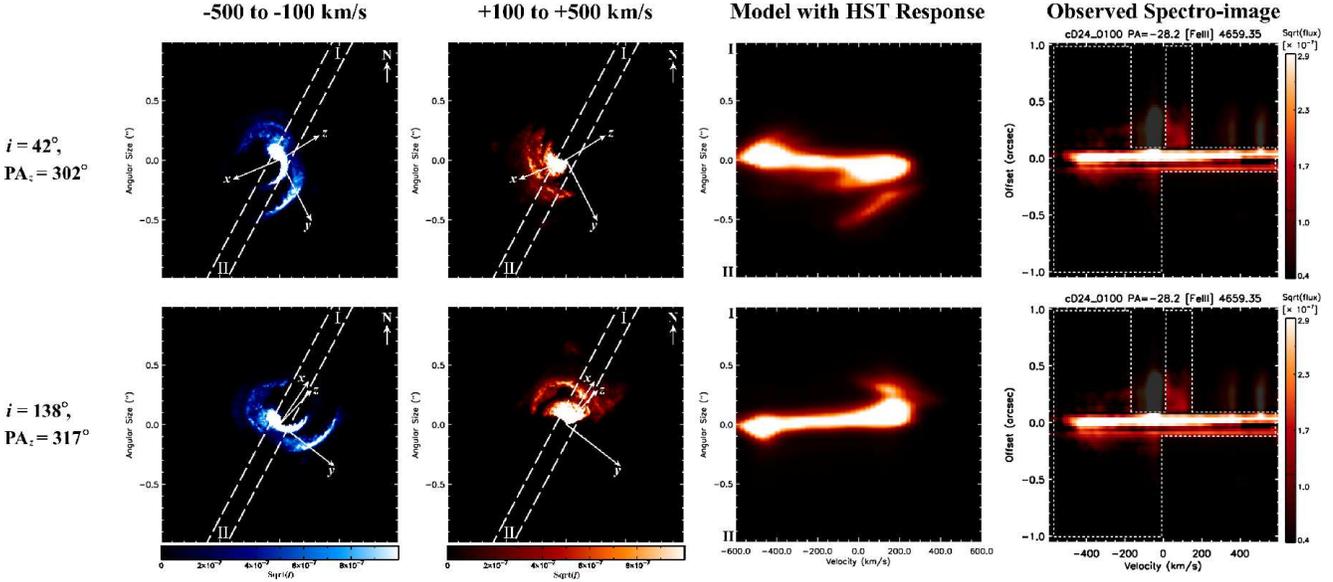}
\caption{Same as Figure \protect\ref{fig9}, but for $\phi = 0.952$, slit $\mathrm{PA} = -28^{\circ}$. Note that the red- and blue-shifted emission in the model spectro-image that assumes $i = 42^{\circ}$ (top row) spatially extends in the wrong directions compared to the observations. \ec's orbital orientation parameters are thus constrained to values of $i \approx 130^{\circ}$ to $145^{\circ}$, $\theta \approx 0^{\circ}$ to $15^{\circ}$, $\mathrm{PA}_{z} \approx 302^{\circ}$ to $327^{\circ}$ (see text).}
\label{fig9_2}
\end{figure*}

\paragraph{Results for Orbital Orientations with a $\mathbf{\btheta \approx 180^{\circ}}$ }

Returning back to Figure~\ref{fig6}, one sees that the spectro-image for $\theta = 180^{\circ}$ fails to match the observations. There is a distinct lack of any spatially-extended, blue-shifted emission. Instead, the extended emission arcs are entirely \emph{red-shifted} due to the quite different $v_{los}$ of the emitting material. When $\theta = 180^{\circ}$, \eca\ is between the observer and \ecb\ during most of the orbit. Therefore, all of the emitting material within \ecb's photoionization zone is on the far side of the system and moving \emph{away from} the observer. The extended forbidden line emission is thus mostly red-shifted. The slit at $\mathrm{PA} = 38^{\circ}$ now primarily samples this red component, producing a spectro-image consisting of entirely red-shifted, spatially-extended arcs, the direct \emph{opposite} of the observations.

Changing the value of $i$ and/or $\mathrm{PA}_{z}$ when $\theta = 180^{\circ}$ does not result in a better match to the observations since most of the material photoionized by \ecb\ and exhibiting forbidden line emission is still moving away from the observer. This is true for all values of $\theta$ near 180$^{\circ}$; none of the model images generated for $135^{\circ} \leq \theta \leq 225^{\circ}$, regardless of the assumed $i$ and $\mathrm{PA}_{z}$, matches the $\phi = 0.976$, $\mathrm{PA} = 38^{\circ}$ observations. Example model spectro-images for these orientations can be found in \citet{madura10}. We therefore find that an orbital orientation that places \ecb\ on the near side of \eca\ at periastron, such as that favored by \citet{falceta05, abraham05b, abrahamfalceta07, kashi07, kashi08, falceta09} and others, is explicitly ruled out.

\paragraph{Results for Orientations with $\mathbf{\btheta \approx 90^{\circ}}$ and $\mathbf{270^{\circ}}$}

Model spectro-images for $\theta = 90^{\circ}$ and $270^{\circ}$ also fail to match the observations. The 2-D images of the modeled [\altion{Fe}{iii}] intensity on the sky in Figure \ref{fig6} show that in both cases, one half of the STIS slit is empty (from center to position II for $\theta = 90^{\circ}$ and from I to the center for $\theta = 270^{\circ}$). Therefore, emission is absent to one side spatially in the spectro-images. Changing the value of $i$ has no effect on this since a change in $i$ at these $\theta$ values is equivalent to a rotation about the $x$ axis in the 2-D intensity images. Instead, a rotation of nearly $90^{\circ}$ in $\mathrm{PA}_{z}$ is needed in order to place emitting material in both halves of the slit. However, such a rotation does not result in a match to the observations. Synthetic spectro-images for $45^{\circ} < \theta < 135^{\circ}$ and $225^{\circ} < \theta < 345^{\circ}$ all suffer from the same problems, namely, either no blue-shifted ring-like emission feature (if $\mathrm{PA}_{z} \approx 132^{\circ}$ or $312^{\circ}$), or arc-like emission to only one side spatially (if $\mathrm{PA}_{z} \approx 42^{\circ}$ or $222^{\circ}$) \citep{madura10}. This is true regardless of the $i$ and $\mathrm{PA}_{z}$ assumed.

\subsubsection{Constraint 2: Variations with Phase at Slit PA = $-28^{\circ}$}

\begin{figure*}
\includegraphics[width=17.75cm]{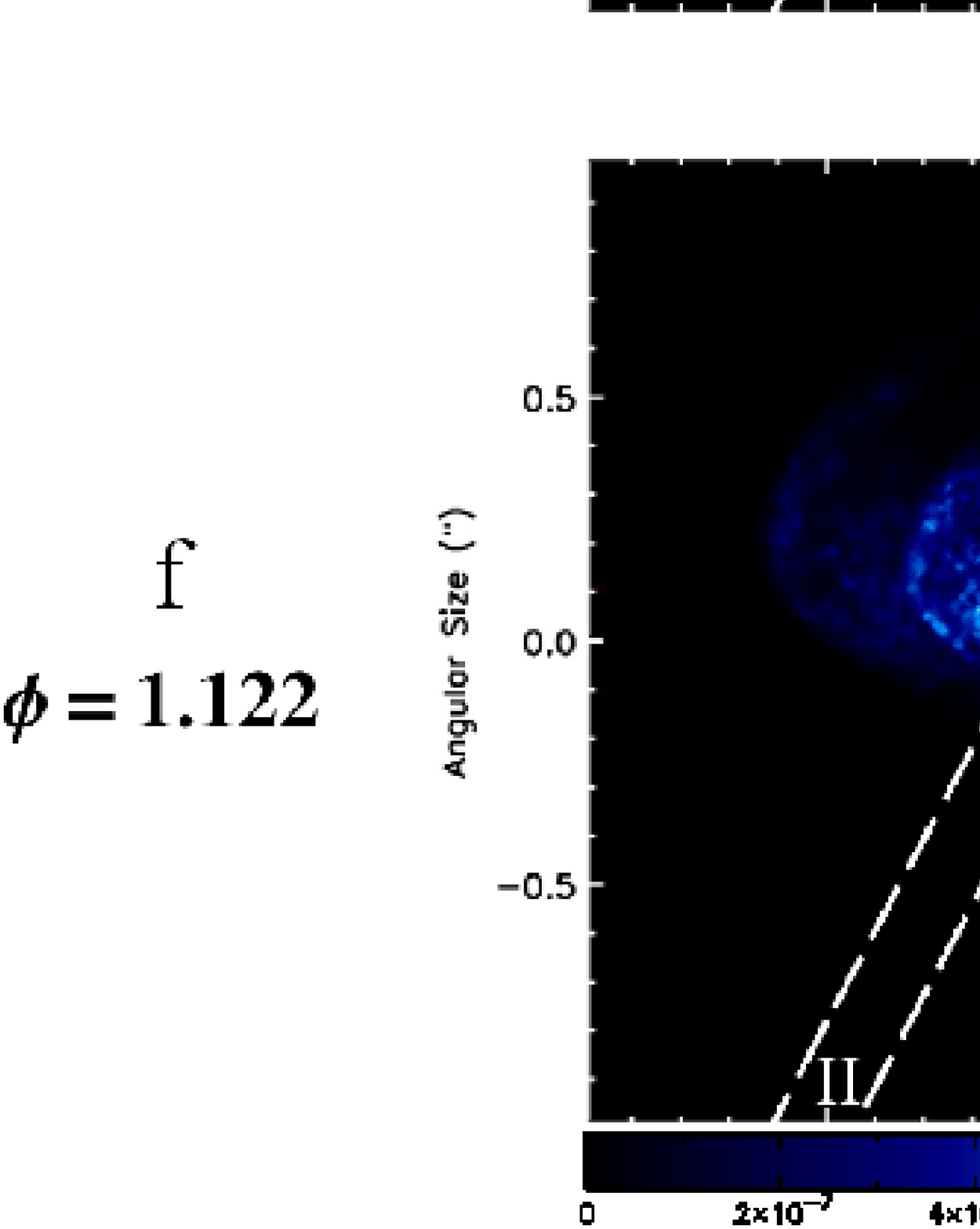}
\caption{Same as Figure \protect\ref{fig9}, but for phases (rows, top to bottom) $\phi = 0.045$, 0.213, 0.407, 0.952, 1.040, and 1.122, with the STIS slit at $\mathrm{PA} = -28^{\circ}$ (as in figure 11 of G09), and assuming $i = 138^{\circ}$, $\theta = 7^{\circ}$, $\mathrm{PA}_{z} = 317^{\circ}$.}
\label{fig12}
\end{figure*}

Using the 3-D dynamical model we generated synthetic spectro-images for a slit $\mathrm{PA} = -28^{\circ}$ at each of the phases in figure 11 of G09, $\phi = 0.045$, 0.213, 0.407, 0.952, 1.040, and 1.122. The main goals were to determine if the binary scenario and 3-D model could explain the phase dependence observed in the high-ionization forbidden lines, and whether the set of $\mathrm{PA} = -28^{\circ}$ observations could resolve the ambiguity between the sets of $i = 35^{\circ} - 50^{\circ}$ and $i = 130^{\circ} - 145^{\circ}$ best-fit orientations found in the above subsection.

Figure \ref{fig9_2} illustrates the results of investigating the two best-fit $i$ regimes. It is clear that synthetic spectro-images for $i \approx 30^{\circ}$ to $50^{\circ}$ \emph{do not} match the observations taken at slit $\mathrm{PA} = -28^{\circ}$ (top row of Figure~\ref{fig9_2}). There are a number of discrepancies, but the most important and obvious is that the blue- and red-shifted emission components in the synthetic spectro-image extend in the wrong spatial directions, with the blue emission stretching to the NW and the red to the SE, the \emph{opposite} of the observations.

In contrast, synthetic images generated for $i \approx 130^{\circ}$ to $150^{\circ}$ (bottom row of Figure~\ref{fig9_2}) \emph{are} capable of matching the observations. The morphology of the $\mathrm{PA} = -28^{\circ}$ observations breaks the degeneracy in $i$ and fully constrains, in 3-D, \ec's orbit. Because red-shifted emission is observed to extend spatially in directions to the NW, the red emission component of [\altion{Fe}{iii}] on the sky must be oriented such that it too stretches NW and falls within the top half of the $\mathrm{PA} = -28^{\circ}$ slit. The synthetic images of the [\altion{Fe}{iii}] intensity on the sky (first two columns of Figures~\ref{fig9_2} and \ref{fig12}) illustrate how the slit at $\mathrm{PA} = -28^{\circ}$ samples the emitting regions of \ec's extended interacting winds in very different directions compared to slit $\mathrm{PA} = 38^{\circ}$. This is why the observed spectro-images are so different between the two slit PAs, even at similar orbital phases. We find that \emph{only} orbital orientations with $i \approx 130^{\circ}$ to $145^{\circ}$, $\theta \approx 0^{\circ}$ to $15^{\circ}$, $\mathrm{PA}_{z} \approx 302^{\circ}$ to $327^{\circ}$ are able to simultaneously match the observations taken at both slit $\mathrm{PA} = 38^{\circ}$ \emph{and} $-28^{\circ}$.

Figure \ref{fig12} shows that the 3-D dynamical model and derived orbital orientation reproduce the overall observed shape and velocity structure of the [\altion{Fe}{iii}] emission at each phase for slit $\mathrm{PA} = -28^{\circ}$. Figure \ref{fig12} further illustrates how \ec's extended interacting winds change with time. The orientation of the blue component does not appear to change, always stretching from NE to SW on the sky. However, the spatial extent does change, growing larger going from periastron to apastron as the wind structures flow outward. Moving from apastron back to periastron, the intensity and spatial extent of the blue component eventually start to decrease as the expanding arcs of primary wind drop in density and mix with the surrounding wind from \ecb. The red-shifted emission changes in a way similar to that of the blue component, but points mainly NW.

\subsubsection{Constraint 3: Observed Variations with $\phi$ and Slit PA}

\begin{table}
\caption{Phases and slit PAs from figures 12 and 13 of G09 modeled in Section 4.3.3 and shown in Figure \protect\ref{fig13}}
\label{tab5.3}
\begin{center}
\begin{tabular}{ccl}
\hline
$\ \ \phi$ & PA & Figure\\ \hline \\
0.601 & $+22^{\circ}$ & \protect\ref{fig13}a\\
0.738 & $-82^{\circ}$ & \protect\ref{fig13}b\\
0.820 & $+69^{\circ}$ & \protect\ref{fig13}c\\
0.930 & $-57^{\circ}$ & \protect\ref{fig13}d\\
0.970 & $+27^{\circ}$ & \protect\ref{fig13}e\\
0.984 & $+62^{\circ}$ & \protect\ref{fig13}f\\
0.995 & $+70^{\circ}$ & \protect\ref{fig13}g\\
1.001 & $+69^{\circ}$ & $\ -$\\
1.013 & $+105^{\circ}$ & $\ -$\\
1.068 & $-142^{\circ}$ & \protect\ref{fig13}h\\ \hline \\
\end{tabular}
\end{center}
\end{table}

Synthetic spectro-images were generated at the ten combinations of phase and slit PA in Table \ref{tab5.3}, assuming the best-fit orientation $i = 138^{\circ}$, $\theta = 7^{\circ}$, and $\mathrm{PA}_{z} = 317^{\circ}$. The goal was to determine how well the 3-D model and derived binary orientation could reproduce observations at a variety of other phases and slit PAs. Model images for $\phi = 1.001$ and $1.013$ are nearly identical to those at $\phi = 0.045$ and $1.040$ in Figure \ref{fig12} and add no new information. We therefore focus this discussion on the phases before and after periastron.

\begin{figure*}
 \subfigure{
   \label{fig13:1}
\includegraphics[width=17.5cm]{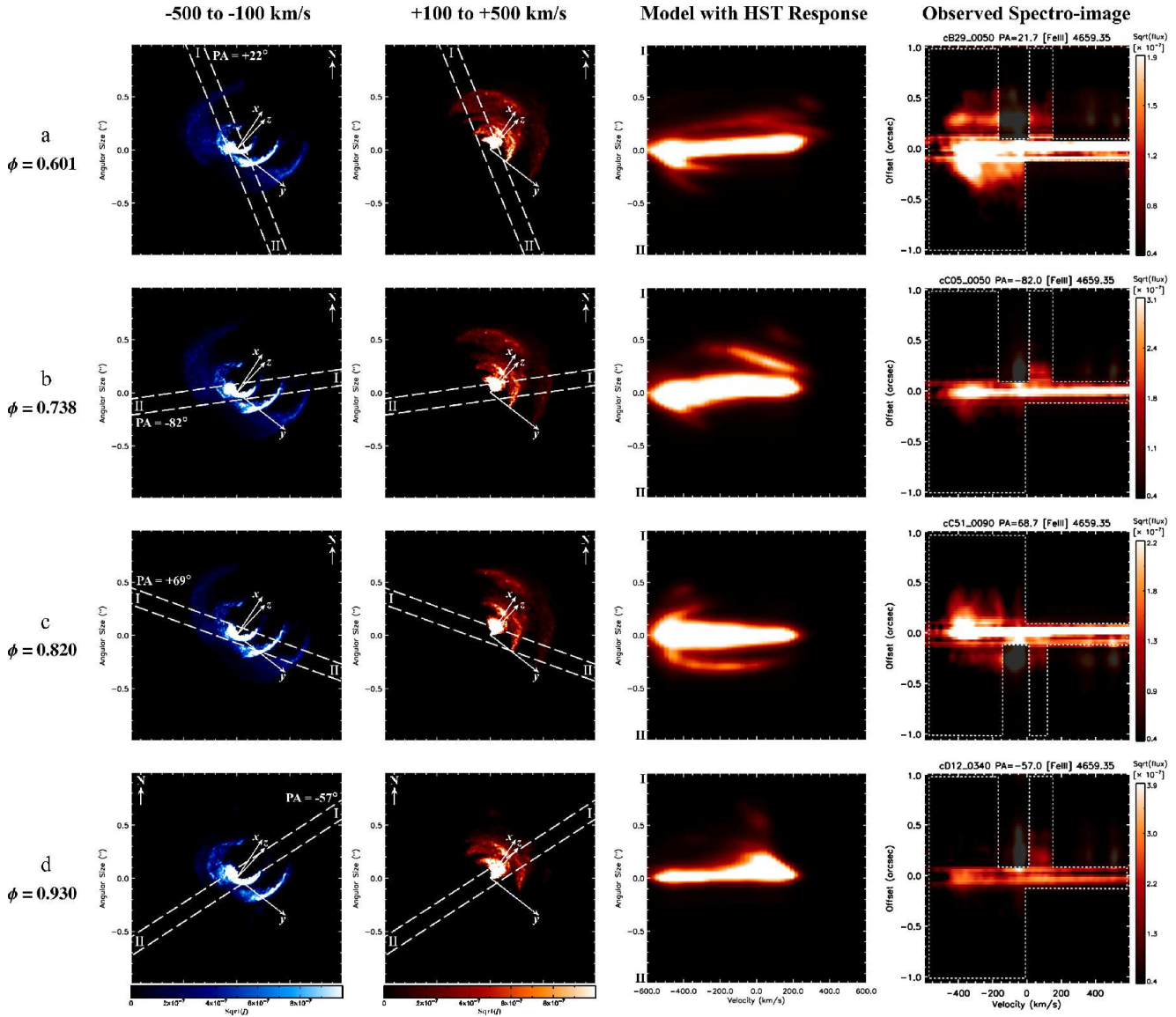}
}
\caption{\textbf{A} Same as Figure \protect\ref{fig12}, but for the observed orbital phases and STIS slit PAs listed in Table \protect\ref{tab5.3}. Phases are (top to bottom): $\phi = 0.601$, 0.738, 0.820, and 0.930. The $0.1''$ wide STIS slit is shown overlaid at the appropriate PA in the 2-D images of the modeled intensity on the sky. Roman numerals I and II indicate the top and bottom of the slit, respectively.}
\label{fig13}
\end{figure*}

Figure \ref{fig13} presents the results. The overall match between synthetic and observed spectro-images is quite good. The model reproduces all of the key spatial features, namely, extended ($> \pm 0.1''$) [\altion{Fe}{iii}] emission that is almost entirely blue-shifted for positive slit PAs, but that is partially red-shifted for negative slit PAs. Observed variations in emission with phase are also reproduced.

\addtocounter{figure}{-1}
\begin{figure*}
 \addtocounter{subfigure}{1}
  \subfigure{
  \label{fig13:2}
\includegraphics[width=17.5cm]{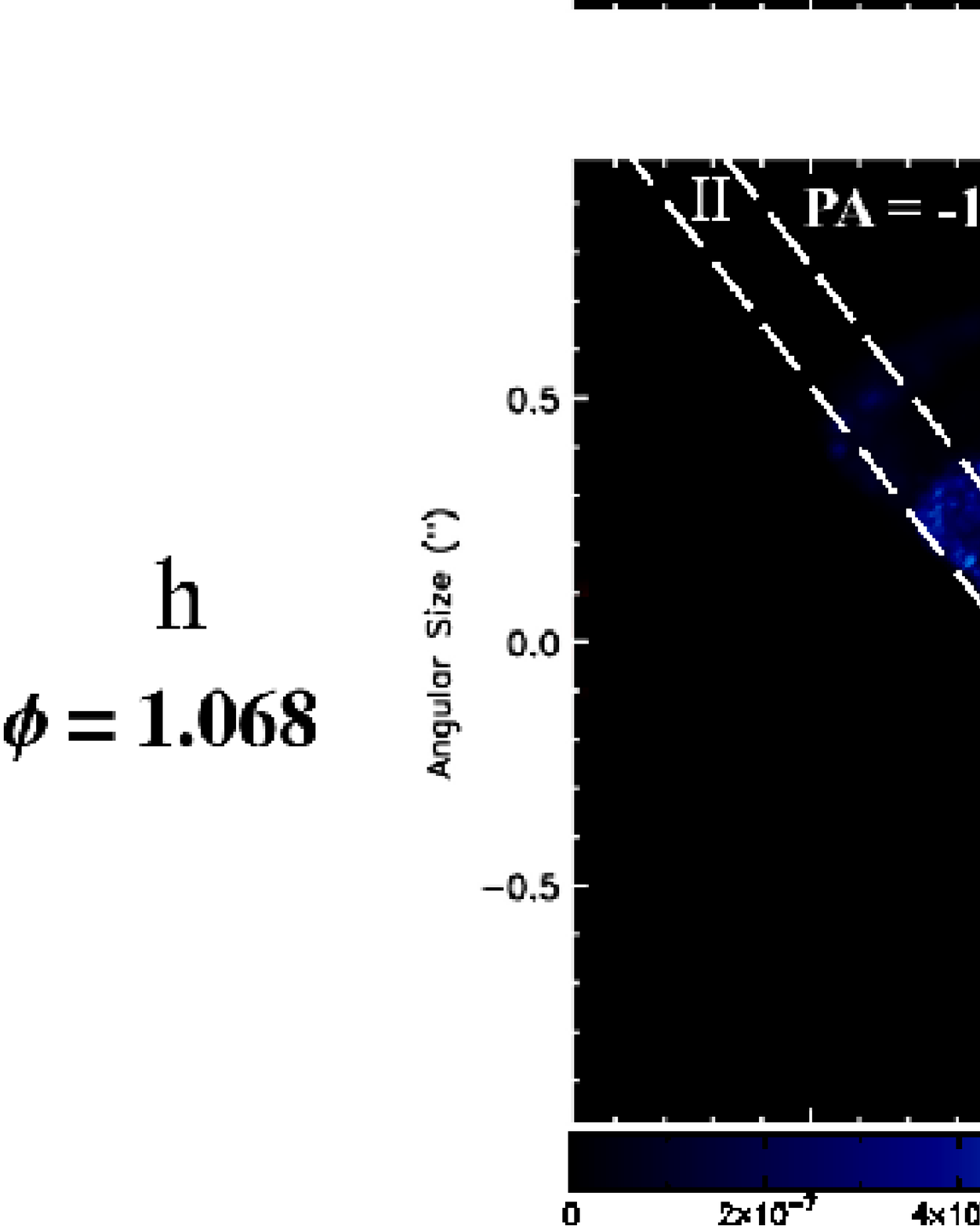}
}
\caption{\textbf{B} Same as Figure \protect\ref{fig13}.A, but at phases (top to bottom) $\phi = 0.970$, 0.984, 0.995, and 1.068.}
\end{figure*}

Model images at $\phi = 0.601$ (\ref{fig13}a) and $0.738$ (\ref{fig13}b) demonstrate how a large change in slit PA results in very different spectro-images. The spatial distributions of the emission on the sky are very similar at these two phases, as expected since the system is near apastron when orbital velocities are their lowest. However, the slit at $\mathrm{PA} = -82^{\circ}$ samples the emission in very different directions compared to $\mathrm{PA} = 22^{\circ}$, producing a spectro-image that resembles those taken at $\mathrm{PA} = -28^{\circ}$ in Figure \ref{fig12}.

The partial ring of emission observed at $\phi = 0.984$, $\mathrm{PA} = 62^{\circ}$ (\ref{fig13}f), is also noteworthy. The match between model and observations is very good, the emission arc to the SW present in both, as well as the much shorter arc to the NE between $\sim -500 \ \mathrm{km} \ \mathrm{s}^{-1}$ and $-350 \ \mathrm{km} \ \mathrm{s}^{-1}$. Two effects appear to be causing the NE arc to be shorter. First, due to their outward expansion, the extended arcs of primary wind have dropped in density and started to mix with the surrounding low-density wind from \ecb. Second, the photoionization of material in directions to the NE by \ecb\ has diminished due to its clockwise orbital motion and gradual embedding in \eca's wind. These lead to a decrease in the amount of extended [\altion{Fe}{iii}] emission to the NE. This, combined with the slit PA, produces a partial arc of blue-shifted emission to the NE in the spectro-image.

By $\phi = 0.995$ (\ref{fig13}g), the extended emission has vanished, implying that the ionizing flux of photons from \ecb\ shuts off sometime between $\phi = 0.984$ and 0.995. At $\phi = 1.068$ (\ref{fig13}h), there is still no observed spatially-extended emission. Yet, the 2-D model images of the intensity on the sky show that the extended [\altion{Fe}{iii}] emission has started to return in directions to the NE. This is visible in the model spectro-image as a bright, blue-shifted bulge that points NE. Unfortunately, the observational data in the central $\pm 0.15''$ is of insufficient quality to tell if such an emission bulge was detected. Nevertheless, the 3-D model predicts that \ecb\ should start to emerge from \eca's dense wind and begin to restore the spatially-extended, high-ionization forbidden line emission at $\phi \approx 1.068$.

\subsubsection{Discrepancies Between the Observations and 3-D Model}

During most of the orbit, the wind of \ecb\ collides with the dense arcs of primary wind on the apastron side of the system where the spatially-extended forbidden line emission forms (see rows b through d of Figure \ref{fig5}). Because the arcs are bordered by a large, low-density wind cavity created earlier by \ecb, they have almost no support against \ecb's high-velocity wind, which is able to drive portions of the arcs into the cavity at velocities just above their outflow speed, the terminal speed of the primary wind. An orientation that places the observer on the apastron side of the system thus results in model spectro-images containing blue-shifted emission at velocities slightly higher than the wind terminal velocity of the primary.

Because the observations do not contain blue-shifted emission at speeds more than $\sim 500 \ \mathrm{km} \ \mathrm{s}^{-1}$, the wind speed(s) of \eca\ and/or \ecb\ used in the SPH simulation may be a bit too large. The launching of the two winds at their terminal velocity in the simulation may also be having a small effect on the results. Better constraints on \ecb's wind terminal speed and improved 3-D simulations with proper driving of the stellar winds should help resolve these issues. Improved modeling coupled with future observations of the extended forbidden emission in \ec's central arcsecond also has the potential to further constrain the wind terminal velocity of \eca.

The main discrepancies between synthetic spectro-images assuming $i = 138^{\circ}$, $\theta = 7^{\circ}$, $\mathrm{PA}_{z} = 317^{\circ}$ and the observations for slit $\mathrm{PA} = -28^{\circ}$ are: (1) the extended blue-shifted component is located $\sim 100 \ \mathrm{km} \ \mathrm{s}^{-1}$ farther to the blue in the model images than in the observations, and (2) the red-shifted component in the model images is $\sim 75 \ \mathrm{km} \ \mathrm{s}^{-1}$ broader than observed. Both issues are likely due to the reasons discussed above concerning the wind terminal velocities of \eca\ and \ecb.

\section{Discussion}

\subsection{The Orientation and Direction of $\boldeta$ Car's Binary Orbit}

There has been much speculation about the binary orientation, and many papers published offering suggestions (see Section $1$). The detailed modeling of Section $4$ tightly constrains the observer's line-of-sight to angles of $\theta \approx 0^{\circ} - 15^{\circ}$ prograde of the semi-major axis on the apastron side of the system, placing \ecb\ behind \eca\ during periastron. Given the uncertainties in the stellar and wind parameters of \ecb\ used in the 3-D simulations, and the assumption that phase zero of the orbital and spectroscopic periods coincide, values of $-15^{\circ} \lesssim \theta \lesssim +30^{\circ}$ may also be possible. A study in the stellar/wind parameters of \ecb, or improved observational constraints, is needed to further refine the value of $\theta$.

\begin{figure}
\begin{center}
\includegraphics[width=8.2cm]{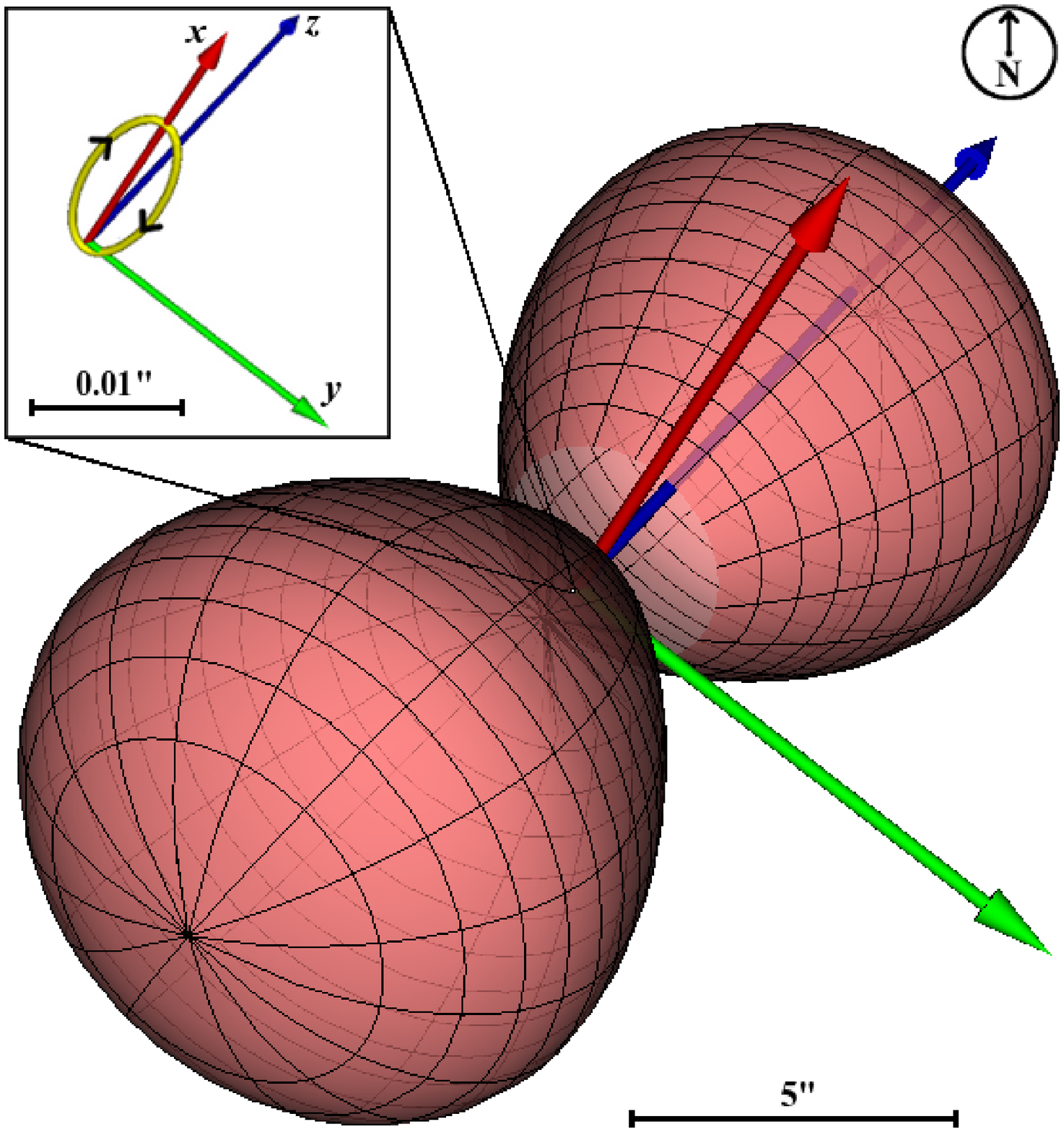}
\end{center}
\caption{Illustration of \ec's binary orbit (inset, yellow) on the sky relative to the Homunculus nebula for a binary orientation with $i = 138^{\circ}$, $\theta = 7^{\circ}$, and $\mathrm{PA}_{z} = 317^{\circ}$, which lies near the center of our best-fit range of orbital parameters. The $+z$ orbital axis (blue) is closely aligned with the Homunculus polar axis in 3-D. \ecb\ orbits clockwise on the sky relative to \eca\ (black arrows in inset), and apastron is on the observer's side of the system. The semi-major axis ($+x$-axis, red) runs from NW to SE on the sky, while the semi-minor axis ($+y$-axis, green) runs from SW to NE. North is up.}
\label{fig14}
\end{figure}

More importantly, the results bound, for the first time, the orientation of \ec's orbital plane, with the orbital axis closely aligned in 3-D with the inferred polar axis of the Homunculus at an $i \approx 130^{\circ} - 145^{\circ}$ and $\mathrm{PA}_{z} \approx 302^{\circ} - 327^{\circ}$. Figure \ref{fig14} illustrates the orientation of \ec's binary orbit on the sky relative to the Homunculus. With $i \approx 138^{\circ}$, $\theta \approx 7^{\circ}$, and $\mathrm{PA}_{z} \approx 317^{\circ}$, the resulting projected orbit on the sky has \ecb\ moving \emph{clockwise} relative to \eca, with \ecb\ approaching \eca\ from the SW prior to periastron, and receding to the NE afterward.

In their recent work modeling \ec's \emph{RXTE} light curve, O08 and P09 investigated values of $i$ in the range $0^{\circ} < i < 90^{\circ}$, and each obtained a best-fit value of $i \approx 42^{\circ}$, which they state is consistent with the orbital axis of the \ec\ binary being aligned with the Homunculus polar axis. However, the inclination angle of the Homunculus, as defined in \citet{davidson01} and \citet{smith06b}, corresponds to the tilt of the polar axis out of the line of sight (figure 5 of Davidson et al. 2001). Since this is the same quantity as the inclination defined for binary orbits, the polar axis of the Homunculus is tilted from the plane of the sky by $\sim 48^{\circ}$. Therefore, an $i \approx 138^{\circ}$ is required in order for the \ec\ orbital axis to be aligned with the Homunculus polar axis, \emph{not} $42^{\circ}$.

The spatially-unresolved nature of the X-ray observations and overall top-bottom symmetry of the problem are what allowed O08 and P09 to reasonably reproduce the \emph{RXTE} light curve using an $i \approx 42^{\circ}$. However, an $i \approx 138^{\circ}$ works as well. Moreover, with either $i$, \emph{any} PA on the sky of the orbital axis produces a good fit to the \emph{RXTE} light curve. Unfortunately, this ambiguity in $i$ and PA of the orbital axis has not been sufficiently discussed in the \ec\ literature, and most have simply assumed that the orbital and Homunculus polar axes are aligned. Yet, knowing the value of $i$ is crucial for proper interpretation of spatially-resolved observational diagnostics and, most importantly, determining the stellar masses.

Taking into account this ambiguity in $i$ inherent to 3-D models of X-ray data, our derived ranges of $i$ and $\theta$ agree very well with those of O08, P09, and P11. They are also consistent with the $i$ and $\omega$ values derived by \citet{groh10b} in their attempts to understand the origin of the high-velocity (up to $-1900 \ \mathrm{km} \ \mathrm{s}^{-1}$) absorption wing detected in \altion{He}{i} $\lambda 10830$ during \ec's 2009 periastron passage. Our binary orientation is in complete agreement with simple models proposed to explain the available radio \citep{duncan03, white05} and \altion{He}{ii} $\lambda 4686$ \citep{teodoro11} observations as well.

Further support for our results and derived orbital orientation comes from recent work by \citet{mehner10}, who mapped the flux of the broad, blue-shifted component of [\altion{Fe}{iii}] $\lambda$4659 and found that most of the emission originates in the inner $0.15''$ region of \ec\ (see their figure 8). The synthetic spectro-images and 2-D spatial maps of [\altion{Fe}{iii}] $\lambda$4659 in Section~$4$ show that most of the emission should indeed originate in the inner $0.15''$. According to our 3-D model, this emission forms in/near the current WWC zone in regions photoionized by \ecb\ and manifests itself in spectro-images as a bright, central streak in the inner $\pm 0.15''$ that spans a wide range of velocities from blue to red. This emission appears as a bright central streak because \emph{HST}/STIS lacks the resolution necessary to resolve the inner WWC region's complex features.

\citet{mehner10} additionally found that the spatial distribution of the blue-shifted component of [\altion{Fe}{iii}] $\lambda$4659 in the central $0.1''$ of \ec\ is not as sharp as a stellar point source and is detectably elongated to the NE and SW (see their figures 8 and 9). Model images displaying the 2-D spatial distribution on the sky of the blue-shifted component of [\altion{Fe}{iii}] $\lambda$4659 (first column of Figures \ref{fig9}, \ref{fig12}, and \ref{fig13}) show that the emission \emph{is} elongated to the NE and SW for our suggested orbital orientation, even in the central $0.1''$. While detailed mapping with \emph{HST}/STIS is needed to verify the results of \citet{mehner10}, they do strongly suggest that our 3-D model and binary orientation are correct.

All other orbital orientations, such as that with $\theta \approx 180^{\circ}$ favored by \citet{falceta05, abraham05b, abrahamfalceta07, kashi07, kashi08, falceta09}; and \citet{kashi09}, produce spectro-images that are in strong disagreement with the $\phi = 0.976$, $\mathrm{PA} = 38^{\circ}$ observations, lacking any spatially-extended, blue-shifted emission and containing significant amounts of extended, red-shifted emission. With \ecb\ the main source of photons capable of producing high-ionization forbidden line emission, there is no realistic situation with $\theta \approx 180^{\circ}$, for any $i$ and $\mathrm{PA}_{z}$, that can produce spectro-images like those observed since there will always be an extended red-shifted component. Hypothesizing some method to absorb or diminish the red-shifted emission does not help since there would then be almost \emph{no} spatially-extended emission detected at slit $\mathrm{PA} = 38^{\circ}$.

There is no known source of ionizing photons in the \ec\ system that permits a $\theta \approx 180^{\circ}$ orientation. Photons generated in the WWC region will not work simply because the WWC zone is on the opposite side of the system during most of the orbit and \eca's dense wind would easily absorb any such photons before they reach the periastron side. Even if some contrived process could produce significant amounts of spatially-extended, blue-shifted emission on the periastron side of the system, this would not result in a match to the observations as the spectro-images would then contain both red- \emph{and} blue-shifted extended emission. There would have to be some way to absorb or diminish the extended, optically-thin red-shifted emission while allowing the blue-shifted emission component to reach the observer. The problem is that extended, red-shifted emission is observed at other slit PAs, even for similar orbital phases. Therefore, the absorption/diminution would have to be restricted to very specific, extended regions on the sky, which seems unrealistic. Given all of these issues, as well as those concerning the observed phase dependence of the emission, an orientation that places apastron on our side of the system seems unavoidable.

\subsubsection{Observational Support from the Weigelt Blobs}

While a detailed review of \ec's spectral variability is beyond the scope of this paper, it now seems well established that observed, phase-dependent, narrow ($\lesssim 50 \ \mathrm{km \ s}^{-1}$) high-ionization forbidden emission lines form in Weigelt blobs B, C, and D \citep{davidson95, zethson01, zethson11, damineli08a, damineli08b, mehner10}. Illumination of the Weigelt blobs by \ecb\ is required for the formation of these narrow lines, which are present during most of \ec's 5.5-year orbit, and fade during periastron passage, gradually returning to their `normal' strength a few months afterward \citep{verner05, mehner10}.

Our proposed orientation and direction for \ec's binary orbit are consistent with, and provide an explanation for, the time variability of the narrow-line emission features seen in the Weigelt blobs. \citet{davidson95, davidsonetal97} found that Weigelt blobs B, C, and D are located within $0.15''$ to $0.3''$ NW of the central stellar source near the Homunculus equatorial plane (see Figure \ref{fig2}). Based on proper motions and observed blue-shifted emission, the Weigelt blobs are located on the same side of \ec\ as the observer and are thought to have been ejected sometime around the smaller eruption of 1890 that also formed the Little Homunculus \citep{davidson97, zethson01, ishibashi03, smith04, nielsen07a, mehner10}. Since the Weigelt blobs are on the same side of the system as the observer, and because \ecb\ is required for the formation of the narrow, high-ionization forbidden lines that form in the blobs during most of the orbit, one easily concludes that \ecb\ must be on the same side of the system as the observer during most of the orbit.

Moreover, the Weigelt blobs are located in the NW quadrant of the system. For our proposed orbital orientation, during most of the orbit, the low-density wind cavity created by \ecb\ is open toward and pointing NW, in the direction of the blobs. The low-density cavity thus provides a path for the ionizing photons from \ecb\ to reach the Weigelt blobs and produce the high-ionization emission. Over the course of the 5.5-year orbit, \ecb\ gradually moves from SE to NW on the sky when going from periastron to apastron, and then from the NW back to the SE when moving from apastron to periastron. The orbital motion of \ecb\ and gradual embedding in the wind of \eca\ close to periastron leads to changes in the spatial extent and direction of the photoionization region, which causes observed variations in the narrow, high-excitation emission in the blobs. During periastron passage, the ionizing flux from \ecb\ shuts off, and the high-excitation emission from the blobs fades. It takes several months after periastron for \ecb\ to restore the large photoionization volume that faces NW, and thus the high-excitation emission in the blobs.

While omitting some of the details, this qualitative picture is consistent with the known behavior of the Weigelt blobs. The proposed clockwise direction of \ecb's orbit is further supported by the images of \citet{smith04}, which show excess UV emission to the SW of \ec\ just before periastron, and to the NE just after. A complete review of all of the available observations of \ec\ and how they support the derived orientation and direction of the binary orbit is obviously not possible here. However, we note that the orientation shown in Figure \ref{fig14} and clockwise motion on the sky of \ecb\ are consistent with all known observations of \ec\ to date.

\subsection{Implications for Theories for the Formation of the Homunculus Nebula and the Nature of $\boldeta_{\mathbf{\mathrm{A}}}$'s Wind}

A variety of models, ranging from the interaction of stellar winds with differing speeds and densities, to binary interactions, and even a binary merger, have been proposed for explaining both the Great Eruption and the bipolar shape of the resulting nebula \citep{smith09}. Observational studies of the Homunculus have firmly established that its polar axis is orientated on the sky at a $\mathrm{PA} \simeq 312^{\circ}$, and that its inclination is $42^{\circ} \pm 1^{\circ}$ \citep{davidson01, smith06b, smith09}. As shown in Figure \ref{fig14}, the orbital axis of the \ec\ binary for our best-fit range of orientation parameters is closely aligned in 3-D with the Homunculus polar axis. Such an alignment has important implications for theories for the formation of the Homunculus and/or the present-day shape of \eca's wind.

\emph{HST}/STIS long-slit spectral observations of the Homunculus by Smith et al. (2003a) indicate that the reflected stellar spectrum over the poles has stronger and broader absorption in H$\alpha$, implying a denser, polar outflow. Very Large Telescope (VLT)/Very Large Telescope Interferometer (VLTI) observations of the optically thick stellar wind of \eca\ by \citet{vanboekel03} and \cite{weigelt07} also indicate that \eca's wind could be prolate in shape and aligned at the same PA on the sky as the Homunculus. Many have interpreted these observational studies as evidence for the current-day wind of \eca\ being prolate, with a bipolar form and orientation similar to the Homunculus. Thus, one explanation offered for both the Homunculus' formation and the nature of the present-day wind is that \eca\ is a rapid rotator (Owocki \& Gayley 1997; Maeder \& Meynet 2000; Dwarkadas \& Owocki 2002; Smith et al. 2003a; Owocki, Gayley, \& Shaviv 2004); the effects of gravity darkening on radiation-driven wind outflows from a rapidly rotating star suggested as a way of explaining both the high polar densities and velocities inferred in \eca's extended wind.

However, using 2-D radiative transfer models, \citet{groh10a} and \citet{groh11} find that the density structure of \eca's wind can be sufficiently disturbed by \ecb, thus strongly affecting the observed UV spectrum, optical hydrogen lines, and mimicking the effects of fast rotation in the interferometric observables. \citet{groh10a} further show that even if \eca\ is a fast rotator, models of the interferometric data are not unique, and both prolate- and oblate-wind models can reproduce the interferometric observations. These prolate- and oblate-wind models additionally suggest that the rotation axis of \eca\ would not be aligned with the Homunculus polar axis. Further complicating the situation are the recent results of \citet{mehner11}, who find no evidence for higher wind velocities at high stellar latitudes in H$\delta$ P Cygni profiles obtained during \ec's normal state (see their Section 5). Together, these new results challenge the idea that the present-day wind of \eca\ is latitudinally-dependent and prolate in shape.

In this context, it is noteworthy that the results in Section 4 match the observations so well, given that the winds from both stars are assumed to be spherical in the 3-D simulations. New 3-D simulations with a latitudinal-dependent wind for \eca, and comparison of the resulting spectro-images to the observations, may provide additional clues as to the nature of \eca's present-day wind. It is important to keep in mind though that these new results do not necessarily mean that \eca's wind was not latitudinally-dependent in the past. Latitudinal-dependent mass-loss from a rapidly-rotating \eca\ could have played a role in the formation of the Homunculus. However, the argument for this possibility should not be made based on the above-mentioned observations of \eca's current-day wind.

If \eca's rotation axis is aligned with the orbital axis, and the star is a rapid rotator, then the extended wind should be prolate and aligned at nearly the same PA as the Homunculus. However, the orbital axis and rotation axis of \eca\ do not have to be aligned. This is another common assumption, with important implications for theories proposed to explain \ec's spectroscopic events \citep{davidson97, smith03a, mehner11}. Future observations using diagnostics that are not significantly affected by \ecb\ are needed to determine whether or not \eca\ is a rapid rotator, and whether its rotation axis is aligned with the orbital axis.

Alignment of the orbital and Homunculus polar axes strongly suggests that binarity played a role in the Great Eruption, and possibly the smaller eruption that later formed the Little Homunculus. A binary merger seems unlikely given the multiple known large eruptions of the system \citep{smith09}. The most likely situation involves some sort of interaction between \eca\ and \ecb\ at periastron. \citet{smith03a} suggested a scenario in which a rotating \eca\ loses mass due to increased angular momentum caused by an interaction with \ecb. Models for the periastron-passage triggering of \ec's massive eruptions, and the requirement of a binary to explain the bipolar shape of the Homunculus, have also been proposed by Soker and colleagues \citep{kashi10, soker04, soker07}.

More recently, \citet{smithfrew10} presented a revised historical light curve for \ec, showing that two 100-day peaks observed in 1838 and 1843 just before the Great Eruption coincide to within weeks of periastron, provided the orbital period then was shorter than the current period by $\sim 5\%$. The beginning of \ec's lesser outburst in 1890 also seems to have occurred around periastron \citep{smithfrew10}. Based on these findings and other considerations, \citet{smith11} proposes that a stellar collision occurred at periastron before and during \ec's Great Eruption, with \ecb\ plunging deeply into the bloated photosphere of \eca. Is close alignment of the orbital axis and polar axis of the Homunculus evidence for such a scenario? Only detailed theoretical modeling can answer this question, but given the now apparent alignment of the two axes, binary interaction scenarios should be seriously considered as possible explanations for \ec's multiple, massive eruptions.

\subsection{Probing Changes in $\boldeta$ Car's Stellar and Wind Parameters via Long-Term Monitoring of the Forbidden Line Emission}

The \ec\ system not only has a centuries-long history of variation, including two major eruptions, it exhibits shorter-term variability with a 5.5-year period \citep{davidson97, humphreys08, smithfrew10}. This implies that \ec's numerous spectroscopic events are also related to periastron passage \citep{damineli08a, damineli08b}. Contrary to many expectations, \ec's 2009.0 spectroscopic event was considerably different than that of earlier cycles \citep[C11;][]{mehner11}. The observed X-ray minimum in 2009 was substantially shorter than the minima in 1998 and 2003, by approximately one month (C10; C11). The emission strength of H$\alpha$ and various X-ray lines also appears to have decreased by factors of order two (C11). Recent \emph{HST} WFPC2 observations by \citet{mehner11} show that the minimum in the UV was much deeper (again by about a factor of two) for the 2009 event than for the 2003 event. While still the subject of debate, the behavior of [\altion{He}{ii}] $\lambda$4686 might have been different in 2009 as well, with a second `outburst' occurring $\sim 30 \ \mathrm{days}$ after the first \citep{teodoro11, mehner11}.

The cause of this sudden change in the behavior of \ec's spectroscopic event is still poorly understood. One proposed explanation is that the mass-loss rate of \eca\ has suddenly decreased by at least a factor of two \citep[C10; C11;][]{mehner11}. However, this is far from confirmed and the exact reasons for such a drop are unknown. The results in this paper suggest a possible way to observationally test the idea that \eca's mass-loss rate has recently changed. All of the \emph{HST}/STIS observations modeled were taken in or before 2004. The model in this paper assumes a mass-loss rate of $10^{-3} \ \mathrm{M}_{\odot} \ \mathrm{yr}^{-1}$ for \eca, which is based on earlier observational and modeling studies (Davidson \& Humphreys 1997; H01; H06). The good match between the synthetic spectro-images and the observations suggests that the mass-loss rate of \eca\ was $\sim 10^{-3} \ \mathrm{M}_{\odot} \ \mathrm{yr}^{-1}$ at the time the observations were taken.

The results in Section 4 though are strongly dependent on the mass-loss rate assumed for \eca\ and the ionizing flux of photons assumed for \ecb. If the ionizing flux of photons from \ecb\ remains constant, but the mass-loss rate of \eca\ drops by a factor of two or more, the size of the photoionization region should increase considerably. Therefore, the spatial extent and flux of the observed high-ionization forbidden lines should also drastically change. The phase dependence of the forbidden emission would likely differ, with the high-ionization emission vanishing at later phases (compared to earlier orbital cycles) when going into periastron, and reappearing at earlier phases afterward. Observed variations with slit PA should be different too, possibly showing longer spatially-extended structures, and even new components at different $v_{los}$. Changes in the wind terminal velocity of \eca\ could be similarly investigated using the maximum observed $v_{los}$ of the spatially-extended emission.

Multi-epoch observations coupled with improved 3-D radiative transfer modeling of the high-ionization forbidden line emission would also help in determining if there is a significant change in \ecb's ionizing flux of photons, mass-loss rate, or wind terminal speed. Constraints on the ionizing flux of photons from \ecb\ could be compared to stellar models for a range of O \citep{martins05} and WR \citep{crowther07} stars, allowing one to obtain a luminosity and temperature for \ecb.

\section{Summary and Conclusions}

A major goal of this paper has been to use the 3-D dynamical model and available \emph{HST}/STIS observations of high-ionization forbidden lines to constrain the absolute 3-D orientation and direction of \ec's binary orbit. The spatially-resolved spectroscopic observations obtained with the \hst/STIS provide the crucial spatial \emph{and} velocity information needed to help accomplish this. Since the forbidden lines do not have an absorption component, they provide a clear view throughout \ec's extended interacting winds. Each high-ionization forbidden line arises \emph{only} in regions where (1) photoionization by \ecb\ can produce the required ion, (2) the density is near the line's critical density, and (3) the material is at the appropriate temperature. Our results are therefore not as dependent on the many complicated details and effects encountered when modeling other forms of emission and/or absorption, such as X-rays or spectral features of \altion{H}{i}, \altion{He}{i}, or other species (H01; H06; Nielsen et al. 2007b; P11).

Synthetic spectro-images of [\altion{Fe}{iii}] emission line structures generated using 3-D SPH simulations of \ec's binary WWC and radiative transfer codes were compared to the available \hst/STIS observations for a variety of orbital phases and STIS slit PAs. The model spectro-images provide important details about the physical mechanisms responsible for the observed high-ionization forbidden line emission, as well as the location and orientation of the observed emitting structures. Below, we summarize our key conclusions and outline the direction of future work.

\begin{enumerate}
  \item Large-scale ($\sim 1600 \ \mathrm{AU}$) 3-D SPH simulations of \ec's binary colliding winds show that during periastron passage, the dense wind of \eca\ flows unimpeded in the direction of system apastron (Figure \ref{fig5}). Following periastron, \ecb's high-velocity wind collides with this primary wind material, creating dense shells that expand outward during \ec's 5.5-year orbital cycle. These shells eventually fragment, forming a pair of dense, spatially-extended `arcs' of primary wind on the apastron side of the system. After several orbital cycles, the arcs drop in density and mix with the surrounding low-density wind material from \ecb.
  \item During most of \ec's orbit, far-UV radiation from \ecb\ photoionizes a significant, spatially-extended region on the apastron side of the system, including portions of the dense arcs of primary wind formed following periastron (Figure \ref{fig5}). Portions of the current WWC region and primary wind just beyond it are also photoionized by \ecb. The density and temperature of the material in these regions are ideal for producing [\altion{Fe}{iii}] emission (Figure \ref{fig5}).
  \item During periastron passage, \ecb\ becomes enveloped in the dense wind of \eca. This significantly reduces the size of the photoionization region, to approximately a few semi-major axes in diameter. Since the large photoionization region has `collapsed', the spatially-extended ($> 0.1''$) high-ionization forbidden line emission vanishes during periastron passage ($\phi \approx 0.99 - 1.07$). It does not reappear until \ecb\ completes periastron passage and restores the extended photoionization region, which can take several months.
  \item Synthetic spectro-images of [\altion{Fe}{iii}] $\lambda$4659 show that most of the [\altion{Fe}{iii}] emission originates in the central $0.15''$, in agreement with figure~8 of \citet{mehner10}. Our 3-D model shows that this emission forms in photoionized material near the current WWC zone, which is not spatially-resolved by \hst/STIS. Spatially-extended ($> 0.15''$) [\altion{Fe}{iii}] emission arises in photoionized portions of the expanding arcs of primary wind formed just after periastron passage.
  \item Only spectro-images generated for lines-of-sight angled $\theta = -15^{\circ}$ to $+30^{\circ}$ prograde of the semi-major axis on the apastron side of the system are able to match the observations taken at $\phi = 0.976$, slit $\mathrm{PA} = 38^{\circ}$. However, there is an ambiguity in the orbital inclination $i$, with models for $i = 35^{\circ}$ to $50^{\circ}$, $\mathrm{PA}_{z} = 272^{\circ}$ to $332^{\circ}$ and $i = 130^{\circ}$ to $145^{\circ}$, $\mathrm{PA}_{z} = 282^{\circ}$ to $342^{\circ}$ both able to produce entirely blue-shifted arcs (Figure \ref{fig9}).
  \item Model spectro-images of multi-phase observations obtained at slit $\mathrm{PA} = -28^{\circ}$ break the above degeneracy in $i$, fully constraining \ec's orbital orientation parameters. Given the uncertainties in some of the stellar/wind parameters used in the 3-D modeling, \emph{our suggested best-fit range of orientation parameters for the \ec\ binary system are} $i \approx 130^{\circ}$ to $145^{\circ}$, $\theta \approx -15^{\circ}$ to $+30^{\circ}$, $\mathrm{PA}_{z} \approx 302^{\circ}$ to $327^{\circ}$. Therefore, \emph{the orbital axis of the \ec\ binary system is closely aligned in 3-D with the Homunculus polar axis, with apastron on the observer's side of the system and \ecb\ orbiting clockwise on the sky relative to \eca} (Figure \ref{fig14}).
  \item All other orbital orientations, including that in which \ecb\ is in front of \eca\ at periastron ($\theta = 180^{\circ}$), are explicitly ruled out as they produce model spectro-images that lack entirely blue-shifted, spatially-extended arcs, containing instead significant amounts of unobserved, spatially-extended red-shifted emission at $\phi = 0.976$, slit $\mathrm{PA} = 38^{\circ}$ (Figure \ref{fig6}).
  \item The 3-D model predicts that the orientation on the sky of the blue-shifted component of [\altion{Fe}{iii}] $\lambda$4659 should not vary much during \ec's orbit, stretching from NE to SW. However, the spatial extent of the emission should grow larger as the system moves from periastron to apastron, and decrease in size close to periastron. The orientation and spatial extent of the red-shifted component should behave in a way similar to that of the blue-shifted component, but pointing NW on the sky.
   \item The apparent alignment of the orbital axis and Homunculus polar axis implies a link between binarity and \ec's numerous massive eruptions. Binary interaction scenarios should be seriously considered as possible explanations for the Great Eruption and formation of the bipolar Homunculus, and possibly the smaller eruption in 1890 that formed the Little Homunculus.
  \item Future detailed 3-D simulations and radiative transfer calculations, together with high-resolution spatial mapping of the high-ionization forbidden line emission with \emph{HST}/STIS, have the potential to further constrain the temperature and luminosity of \ecb. The strong dependence of the forbidden line emission on the mass-loss rate of \eca\ and ionizing flux of photons from \ecb\ may provide clues as to the nature of \ec's long-term variability, specifically, the cycle-to-cycle variations of the spectroscopic events.
\end{enumerate}

The 3-D dynamical modeling in this paper and the available \emph{HST}/STIS spectral observations have increased our understanding of the \ec\ system. Much work remains, however. Efforts are underway to improve both the 3-D hydrodynamical modeling and 3-D radiative transfer approach. New 3-D simulations that include radiative cooling, radiation-driven stellar winds, gravity, and effects like radiative inhibition and braking are on the horizon. Preliminary results of detailed 3-D radiative transfer calculations using \texttt{SimpleX} \citep{paardekooper10} are promising as well. With these new tools at our disposal, it should be possible to further constrain the stellar, wind, and orbital parameters of \ec, setting the stage for orbital modeling to determine the stellar masses.


\section*{Acknowledgements}
T.~I.~M. and C.~M.~P.~R. were supported through the NASA Graduate Student Researchers Program. T.~I.~M. and J.~H.~G. also thank the Max Planck Society for financial support for this work. All SPH simulations were performed on NASA Advanced Supercomputing (NAS) facilities using NASA High End Computing (HEC) time. S.~P.~O. acknowledges partial support from NASA ATP grant NNK11AC40G. T.~R.~G. acknowledges the hospitality of the MPIfR during the completion of this paper.

\bsp

\label{lastpage}


\begin{thebibliography}{99}
\bibitem[Abdo et al.(2010)]{abdo10} Abdo, A.~A., et al.\ 2010, ApJ, 723, 649
\bibitem[Abraham et al.(2005a)]{abraham05a} Abraham, Z., Falceta-Gon{\c c}alves, D., Dominici, T.~P., Nyman, L.-{\AA}., Durouchoux, P., McAuliffe, F., Caproni, A., and Jatenco-Pereira, V.\ 2005a, A\&A, 437, 977
\bibitem[Abraham et al.(2005b)]{abraham05b} Abraham Z., Falceta-Gon\c{c}alves D., Dominici T., Caproni A., and Jatenco-Pereira V.\ 2005b, MNRAS, 364, 922
\bibitem[Abraham \& Falceta-Gon\c{c}alves(2007)]{abrahamfalceta07} Abraham Z. and Falceta-Gon\c{c}alves D.\ 2007, MNRAS, 378, 309
\bibitem[Abraham \& Damineli(1999)]{abraham99} Abraham, Z. and Damineli, A.\ 1999, Eta Carinae at The Millennium, 179, 263
\bibitem[Bate \& Bonnell(1997)]{bate97} Bate, M.~R. and Bonnell, I.~A.\ 1997, MNRAS, 285, 33
\bibitem[Bate et al.(1995)]{bate95} Bate, M.~R., Bonnell, I.~A., and Price, N.~M.\ 1995, MNRAS, 277, 362
\bibitem[Benz et al.(1990)]{benz90} Benz, W., Cameron, A.~G.~W., Press, W.~H., and Bowers, R.~L.\ 1990, ApJ, 348, 647
\bibitem[Bryans et al.(2006)]{bryans06} Bryans, P., Badnell, N.~R., Gorczyca, T.~W., Laming, J.~M., Mitthumsiri, W., and Savin, D.~W.\ 2006, ApJS, 167, 343
\bibitem[Corcoran(2011)]{corcoran11} Corcoran, M.~F.\ 2011, Bulletin de la Societe Royale des Sciences de Liege, 80, 578 (C11)
\bibitem[Corcoran et al.(2010)]{corcoran10} Corcoran, M.~F., Hamaguchi, K., Pittard, J.~M., Russell, C.~M.~P., Owocki, S.~P., Parkin,E.~R., and Okazaki, A.\ 2010, ApJ, 725, 1528 (C10)
\bibitem[Corcoran(2005)]{corcoran05} Corcoran, M.F.\ 2005, ApJ, 129, 2018
\bibitem[Corcoran et al.(2004)]{corcoran04} Corcoran, M. F., et al.\ 2004, ApJ, 613, 381
\bibitem[Corcoran et al.(2000)]{corcoran00} Corcoran, M. F., Fredericks, A. C., Petre, R., Swank, J. H., and Drake, S. A.\ 2000, ApJ, 545, 420
\bibitem[Cox et al.(1995)]{cox95} Cox, P., Mezger, P.~G., Sievers, A., Najarro, F., Bronfman, L., Kreysa, E., and Haslam, G.\ 1995, A\&A, 297, 168
\bibitem[Crowther(2007)]{crowther07} Crowther, P.~A.\ 2007, ARA\&A, 45, 177
\bibitem[Damineli et al.(2008a)]{damineli08a} Damineli, A. et al.\ 2008a, MNRAS, 384, 1649
\bibitem[Damineli et al.(2008b)]{damineli08b} Damineli, A. et al.\ 2008b, MNRAS, 386, 2330
\bibitem[Damineli et al.(2000)]{damineli00} Damineli, A., Kaufer, A., Wolf, B., Stahl, O., Lopes, D.~F., and de Ara{\'u}jo, F.~X.\ 2000, ApJL, 528, L101
\bibitem[Damineli(1996)]{damineli96} Damineli, A.\ 1996, ApJL, 460, L49
\bibitem[Davidson(2005)]{davidson05} Davidson, K.\ 2005, The Fate of the Most Massive Stars, 332, 101
\bibitem[Davidson et al.(2001)]{davidson01} Davidson, K., Smith, N., Gull, T.~R., Ishibashi, K., and Hillier, D.~J.\ 2001, AJ, 121, 1569
\bibitem[Davidson(1999)]{davidson99} Davidson, K.\ 1999, Eta Carinae at The Millennium, 179, 304
\bibitem[Davidson et al.(1999)]{davidsonetal99} Davidson, K., Ishibashi, K., Gull, T.~R., and Humphreys, R.~M.\ 1999, Eta Carinae at The Millennium, 179, 227
\bibitem[Davidson(1997)]{davidsonnewa97} Davidson, K.\ 1997, NewA, 2, 387
\bibitem[Davidson et al.(1997)]{davidsonetal97} Davidson, K. et al.\ 1997, AJ, 113, 335
\bibitem[Davidson \& Humphreys(1997)]{davidson97} Davidson, K. and Humphreys, R.M.\ 1997, ARA\&A, 35, 1
\bibitem[Davidson et al.(1995)]{davidson95} Davidson K., Ebbets D., Weigelt G., Humphreys R. M., Hajian A. R., Walborn N. R., and Rosa M.\ 1995, ApJ, 109, 1784
\bibitem[Dopita \& Sutherland(2003)]{dopita03} Dopita, M. A. and Sutherland, R. S.\ 2003, Diffuse Matter in the Universe, Springer: Heidelberg
\bibitem[Duncan \& White(2003)]{duncan03} Duncan, R.~A. and White, S.~M.\ 2003, MNRAS, 338, 425
\bibitem[Dwarkadas \& Owocki(2002)]{dwark02} Dwarkadas, V. and Owocki, S.P.\ 2002, ApJ, 581, 1337
\bibitem[Falceta-Gon\c{c}alves \& Abraham(2009)]{falceta09} Falceta-Gon\c{c}alves D. and Abraham, Z.\ 2009, MNRAS, 399, 1441
\bibitem[Falceta-Gon\c{c}alves et al.(2005)]{falceta05} Falceta-Gon\c{c}alves D., Jatenco-Pereira V., and Abraham Z.\ 2005, MNRAS, 357, 895
\bibitem[Feast et al.(2001)]{feast01} Feast M., Whitelock P., and Marang F.\ 2001, MNRAS, 322, 741
\bibitem[Fern{\'a}ndez-Laj{\'u}s et al.(2010)]{eduardo10} Fern{\'a}ndez-Laj{\'u}s, E. et al.\ 2010, New Astronomy, 15, 108
\bibitem[Groh(2011)]{groh11} Groh, J.\ 2011, Bulletin de la Societe Royale des Sciences de Liege, 80, 590
\bibitem[Groh et al.(2010b)]{groh10b} Groh, J.~H., et al.\ 2010b, A\&A, 517, A9
\bibitem[Groh et al.(2010a)]{groh10a} Groh, J.~H., Madura, T.~I., Owocki, S.~P., Hillier, D.~J., \& Weigelt, G.\ 2010a, ApJL, 716, L223
\bibitem[Gull et al.(2009)]{gull09} Gull, T.R., et al.\ 2009, MNRAS, 396, 1308 (G09)
\bibitem[Hamaguchi et al.(2007)]{hamaguchi07} Hamaguchi, K. et al.\ 2007, ApJ, 663, 522
\bibitem[Hartman(2003)]{hartman03} Hartman, H.\ 2003, Ph.D. Dissertation, Lund Observatory, Lund University, Sweden
\bibitem[Henley et al.(2008)]{henley08} Henley, D.~B., Corcoran, M.~F., Pittard, J.~M., Stevens, I.~R., Hamaguchi, K., and Gull, T.~R.\ 2008, ApJ, 680, 705
\bibitem[Henley(2005)]{Henley05} Henley, D.B.\ 2005, Ph.D. Dissertation, University of Birmingham
\bibitem[Hillier et al.(2006)]{hillier06} Hillier, D.~J. et al.\ 2006, ApJ, 642, 1098 (H06)
\bibitem[Hillier et al.(2001)]{hillier01} Hillier, D.~J., Davidson, K., Ishibashi, K., and Gull, T.\ 2001, ApJ, 553, 837 (H01)
\bibitem[Hillier \& Miller(1998)]{hillier98} Hillier, D. J. and Miller, D. L.\ 1998, ApJ, 496, 407
\bibitem[Hillier \& Allen(1992)]{hillier92} Hillier, D.J. and Allen, D.A.\ 1992, A\&A, 262, 153
\bibitem[Hillier(1988)]{hillier88} Hillier D. J.\ 1988, ApJ, 327, 822
\bibitem[Humphreys et al.(2008)]{humphreys08} Humphreys, R.~M.,
Davidson, K., and Koppelman, M.\ 2008, AJ, 135, 1249
\bibitem[Iben(1999)]{iben99} Iben, I., Jr.\ 1999, Eta Carinae at The Millennium, 179, 367
\bibitem[Ignace et al.(2009)]{ignace09} Ignace, R., Bessey, R., and Price, C.~S.\ 2009, MNRAS, 395, 962
\bibitem[Ignace \& Brimeyer(2006)]{ignace06} Ignace, R. and Brimeyer, A.\ 2006, MNRAS, 371, 343
\bibitem[Ishibashi et al.(2003)]{ishibashi03} Ishibashi, K., et al.\ 2003, AJ, 125, 3222
\bibitem[Ishibashi et al.(1999)]{ishibashi99} Ishibashi, K., Corcoran, M. F., Davidson, K., Swank, J. H., Petre, R., Drake, S. A., Damineli, A., and White, S.\ 1999, ApJ, 524, 983
\bibitem[Kashi \& Soker(2010)]{kashi10} Kashi, A. and Soker, N.\ 2010, ApJ, 723, 602
\bibitem[Kashi \& Soker(2009)]{kashi09} Kashi, A. and Soker, N.\ 2009, MNRAS, 397, 1426
\bibitem[Kashi \& Soker(2008)]{kashi08} Kashi, A. and Soker, N.\ 2008, MNRAS, 390, 1751
\bibitem[Kashi \& Soker(2007)]{kashi07} Kashi, A. and Soker, N.\ 2007, NewA, 12, 590
\bibitem[Krist \& Hook(1999)]{krist99} Krist, J. and Hook, R.\ 1999, The Tiny Tim User's Guide (Baltimore: STScI)
\bibitem[Madura(2010)]{madura10} Madura, T.~I.\ 2010, Ph.D. Dissertation, University of Delaware
\bibitem[Maeder \& Meynet(2000)]{maeder00} Maeder, A. and Meynet, G.\ 2000, ARA\&A, 38, 143
\bibitem[Martins et al.(2005)]{martins05} Martins, F., Schaerer, D., and Hillier, D.~J.\ 2005, A\&A, 436, 1049
\bibitem[Mehner et al.(2011)]{mehner11} Mehner, A., Davidson, K., Martin, J.~C., Humphreys, R.~M., Ishibashi, K., and Ferland, G.~J.\ 2011, arXiv:1106.5869
\bibitem[Mehner et al.(2010)]{mehner10} Mehner, A., Davidson, K., Ferland, G.~J., and Humphreys, R.~M.\ 2010, ApJ, 710, 729
\bibitem[Mihalas(1978)]{mihalas78} Mihalas, D.\ 1978, Stellar Atmospheres (New York: W. H. Freeman)
\bibitem[Monaghan(2005)]{monaghan05} Monaghan, J.~J.\ 2005, Reports on Progress in Physics, 68, 1703
\bibitem[Nahar(1997)]{nahar97} Nahar, S.~N.\ 1997, PhRvA, 55, 1980
\bibitem[Nahar \& Pradhan(1996)]{nahar96} Nahar, S.~N. and Pradhan, A.~K.\ 1996, A\&AS, 119, 509
\bibitem[Nielsen et al.(2007a)]{nielsen07a} Nielsen, K.~E., Ivarsson, S., and Gull, T.~R.\ 2007a, ApJS, 168, 289
\bibitem[Nielsen et al.(2007b)]{nielsen07b} Nielsen, K.~E., Corcoran, M.~F., Gull, T.~R., Hillier, D.~J., Hamaguchi, K., Ivarsson, S., and Lindler, D.~J.\ 2007b, ApJ, 660, 669
\bibitem[Nussbaumer \& Vogel(1987)]{nussbaumer87} Nussbaumer, H. and Vogel, M.\ 1987, A\&A, 182, 51
\bibitem[Okazaki et al.(2008)]{okazaki08} Okazaki, A. T., Owocki, S. P., Russell, C. M. P., and Corcoran, M. F.\ 2008, MNRAS, 388, L39 (O08)
\bibitem[Osterbrock(1989)]{osterbrock89} Osterbrock, D. E.\ 1989, Astrophysics of Gaseous Nebulae and Active Galactic Nuclei, University Science Books, California
\bibitem[Owocki et al.(2004)]{owocki04} Owocki, S.~P., Gayley, K.~G., and Shaviv, N.~J.\ 2004, ApJ, 616, 525
\bibitem[Owocki \& Gayley(1995)]{owocki95} Owocki, S. P. and Gayley, K. G.\ 1995, ApJ, 454, L145
\bibitem[Paardekooper(2010)]{paardekooper10} Paardekooper, J-P.\ 2010, Ph.D. Dissertation, Leiden Observatory, Leiden University, Netherlands
\bibitem[Parkin et al.(2011)]{parkin10} Parkin, E.~R., Pittard, J.~M., Corcoran, M.~F., and Hamaguchi, K.\ 2011, ApJ, 726, 105 (P11)
\bibitem[Parkin et al.(2009)]{parkin09} Parkin, E. R., Pittard, J. M., Corcoran, M. F., Hamaguchi, K., and Stevens, I. R.\ 2009, MNRAS, 394, 1758 (P09)
\bibitem[Pittard \& Corcoran(2003)]{pittard03} Pittard, J.M. and Corcoran, M.F.\ 2003, RMxAC, 15, 81
\bibitem[Pittard \& Corcoran(2002)]{pittard02} Pittard, J.M. and Corcoran, M.F.\ 2002, A\&A, 383, 636 (PC02)
\bibitem[Pittard(2000)]{pittard00} Pittard, J.M.\ 2000, Ph.D. Dissertation, University of Birmingham
\bibitem[Pittard(1998)]{pittard98} Pittard, J. M.\ 1998, MNRAS, 300, 479
\bibitem[Price(2007)]{price07} Price, D.~J.\ 2007, Publications of the Astronomical Society of Australia, 24, 159
\bibitem[Price(2004)]{price04} Price, D.~J.\ 2004, Ph.D. Dissertation, Institute of Astronomy \& Churchill College, University of Cambridge
\bibitem[Richardson et al.(2010)]{richardson10} Richardson, N.~D., Gies, D.~R., Henry, T.~J., Fern{\'a}ndez-Laj{\'u}s, E., and Okazaki, A.~T.\ 2010, AJ, 139, 1534
\bibitem[Smith \& Frew(2011)]{smithfrew10} Smith, N. and Frew, D.~J.\ 2011, MNRAS, 415, 2009
\bibitem[Smith(2011)]{smith11} Smith, N.\ 2011, MNRAS, 415, 2020
\bibitem[Smith(2009)]{smith09} Smith, N.\ 2009, arXiv:0906.2204
\bibitem[Smith(2006b)]{smith06b} Smith, N.\ 2006b, ApJ, 644, 1151
\bibitem[Smith \& Owocki(2006)]{smithowocki06} Smith, N. and Owocki, S.\ 2006, ApJ, 645, L45
\bibitem[Smith et al.(2004)]{smith04} Smith, N. et al.\ 2004, ApJ, 605, 405
\bibitem[Smith et al.(2003a)]{smith03a} Smith, N., Davidson, K., Gull, T.~R., Ishibashi, K., and Hillier, D.~J.\ 2003a, ApJ, 586, 432
\bibitem[Smith et al.(2003b)]{smith03b} Smith, N., Gehrz, R.~D., Hinz, P.~M., Hoffmann, W.~F., Hora, J.~L., Mamajek, E.~E., and Meyer, M.~R.\ 2003b, AJ, 125, 1458
\bibitem[Soker(2007)]{soker07} Soker, N.\ 2007, ApJ, 661, 490
\bibitem[Soker(2004)]{soker04} Soker, N.\ 2004, ApJ, 612, 1060
\bibitem[Tavani et al.(2009)]{tavani09} Tavani, M., et al.\ 2009, ApJL, 698, L142
\bibitem[Teodoro et al.(2011)]{teodoro11} Teodoro, M., et al.\ 2011, arXiv:1104.2276
\bibitem[van Boekel et al.(2003)]{vanboekel03} van Boekel, R., et al.\ 2003, A\&A, 410, L37
\bibitem[van Genderen et al.(2006)]{vangenderen06} van Genderen, A.~M., Sterken, C., Allen, W.~H., and Walker, W.~S.~G.\ 2006, Journal of Astronomical Data, 12, 3
\bibitem[Verner et al.(2005)]{verner05} Verner, E., Bruhweiler, F., and Gull, T.\ 2005, ApJ, 624, 973
\bibitem[Vishniac(1983)]{vishniac83} Vishniac, E.T.\ 1983, ApJ, 274, 152
\bibitem[Weigelt et al.(2007)]{weigelt07} Weigelt, G., et al.\ 2007, A\&A, 464, 87
\bibitem[Weigelt \& Ebersberger(1986)]{weigelt86} Weigelt, G. and Ebersberger, J.\ 1986, A\&A, 163, L5
\bibitem[White et al.(2005)]{white05} White, S.~M., Duncan, R.~A., Chapman, J.~M., and Koribalski, B.\ 2005, The Fate of the Most Massive Stars, 332, 126
\bibitem[Whitelock et al.(2004)]{whitelock04} Whitelock, P.~A., Feast, M.~W., Marang, F., and Breedt, E.\ 2004, MNRAS, 352, 447
\bibitem[Williams(2008)]{williams08} Williams, P.~M.\ 2008, Revista Mexicana de Astronomia y Astrofisica Conference Series, 33, 71
\bibitem[W\"{u}nsch et al.(2010)]{wunsch10} W\"{u}nsch, R., Dale, J.E., Palous, J., and Whitworth, A.P.\ 2010, MNRAS, 407, 1963
\bibitem[Zanella et al.(1984)]{zanella84} Zanella, R., Wolf, B., and Stahl, O.\ 1984, A\&A, 137, 79
\bibitem[Zethson et al.(2011)]{zethson11} Zethson, T., Johansson, S., Hartman, H., and Gull, T.R.\ 2011, A\&A, submitted
\bibitem[Zethson(2001)]{zethson01} Zethson, T.\ 2001, Ph.D. Dissertation, Lunds Universitet, Sweden
\bibitem[Zhang(1996)]{zhang96} Zhang, H.\ 1996, A\&AS, 119, 523
\end{thebibliography}
\end{document}